\documentclass[12pt]{article}
\usepackage{graphicx}
\usepackage[margin=1.25in]{geometry}
\usepackage{color}
\usepackage{url}
\usepackage[colorlinks = true,
            linkcolor = blue,
            urlcolor  = blue,
            citecolor = blue,
            anchorcolor = blue]{hyperref}

\newcommand\pubnumber{ILC-NOTE-2015-067\\ 
DESY 15-094\\  KEK Preprint 2015-16\\  LAL 15-188 \\ MPP-2015-120 \\
SLAC--PUB--16302}
\newcommand\pubdate{June, 2015}

\def\SLAC{SLAC,
    Stanford University, Menlo Park, CA 94025, USA}

\def\Title#1{\begin{center} {\Large #1 } \end{center}}
\def\Author#1{\begin{center}{ \sc #1} \end{center}}

\newcommand\pubblock{\rightline{\begin{tabular}{l} \pubnumber\\
         \pubdate \end{tabular}}}
\newenvironment{Abstract}{\begin{quotation} \begin{center}
                       ABSTRACT
     \end{center}\bigskip  }{\end{quotation}}

\def\Acknowledgements{\bigskip  \bigskip \begin{center} 
             \bf ACKNOWLEDGEMENTS \end{center}}
\def\kek{High Energy Accelerator Research Organization (KEK), Tsukuba,
  Ibaraki, JAPAN  }
\def\ifae{ICREA at IFAE, Univesitat Aut\'onoma de Barcelona, E-08193
  Bellaterra, SPAIN   }
\def\Tsinghua{Center for High Energy Physics, Tsinghua University, Beijing, CHINA}
\def\Toyama{Department of Physics, University of Toyama, Toyama 930-8555, JAPAN}
\def\Seoul{Department of Physics and Astronomy, Seoul National
  University, Seoul 151-747,  \\ \hskip 0.4in   KOREA}
\def\DESY{DESY, Notkestrasse 85, 22607 Hamburg, GERMANY}
\def\Cornell{Laboratory for Elementary Particle Physics, Cornell
  University, Ithaca, NY 14853, \\ \hskip 0.4in USA  }
\def\Orsay{LAL, Centre Scientifique d'Orsay, Universit\'e Paris-Sud, 
  F-91898 Orsay CEDEX, \\ \hskip 0.4in FRANCE }
\def\MPP{Max-Planck-Institut f\"ür Physik, F\"öhringer Ring 6, 80805 Munich, GERMANY}
\def\Tokyo{ICEPP, University of Tokyo, Hongo, Bunkyo-ku, Tokyo,
  113-0033, JAPAN}
\def\UTA{Department of Physics, University of Texas, Arlington, TX 76019, USA}
\def\Michigan{Michigan Center for Theoretical Physics, University of
  Michigan, Ann Arbor, \\ \hskip 0.4in MI 48109, USA}
\def\Berkeley{Department of Physics, University of California,
  Berkeley, CA 94720, USA}
\def\LBL{Theoretical Physics Group, Lawrence Berkeley National
  Laboratory, Berkeley, \\  \hskip 0.4in   CA 94720, USA}
\def\IPMU{Kavli Institute for the Physics and Mathematics of the
  Universe, \\  \hskip 0.4in   University of Tokyo, Kashiwa 277-8583, JAPAN}
\def\Tohoku{Department of Physics, Tohoku University, Sendai, Miyagi 980-8578, JAPAN}

\def\beq{\begin{equation}}
\def\eeq#1{\label{#1}\end{equation}}
\def\eeqn{\end{equation}}

\newenvironment{Eqnarray}%
   {\arraycolsep 0.14em\begin{eqnarray}}{\end{eqnarray}}
\def\beqa{\begin{Eqnarray}}
\def\eeqa#1{\label{#1}\end{Eqnarray}}
\def\eeqan{\end{Eqnarray}}

\def\leqn#1{(\ref{#1})}

\let\bar=\overbar

\def\lsim{\mathrel{\raise.3ex\hbox{$<$\kern-.75em\lower1ex\hbox{$\sim$}}}}
\def\gsim{\mathrel{\raise.3ex\hbox{$>$\kern-.75em\lower1ex\hbox{$\sim$}}}}

\def\del{\partial}
\def\Dslash{\not{\hbox{\kern-4pt $D$}}}
\def\dslash{\not{\hbox{\kern-2pt $\del$}}}

\def\ee{e^+e^-}

\def\msb{{\bar{\scriptsize M \kern -1pt S}}}

\def\drb{{\bar{\scriptsize D \kern -1pt R}}}

\def\s#1{\widetilde{#1}}

\makeatletter
\def\section{\@startsection{section}{0}{\z@}{5.5ex plus .5ex minus
 1.5ex}{2.3ex plus .2ex}{\large\bf}}
\def\subsection{\@startsection{subsection}{1}{\z@}{3.5ex plus .5ex minus
 1.5ex}{1.3ex plus .2ex}{\normalsize\bf}}
\def\subsubsection{\@startsection{subsubsection}{2}{\z@}{-3.5ex plus
-1ex minus  -.2ex}{2.3ex plus .2ex}{\normalsize\sl}}

\renewcommand{\@makecaption}[2]{%
   \vskip 10pt
   \setbox\@tempboxa\hbox{\small #1: #2}
   \ifdim \wd\@tempboxa >\hsize     
       \small #1: #2\par          
     \else                        
       \hbox to\hsize{\hfil\box\@tempboxa\hfil}
   \fi}

 \def\citenum#1{{\def\@cite##1##2{##1}\cite{#1}}}
 
\newcount\@tempcntc
\def\@citex[#1]#2{\if@filesw\immediate\write\@auxout{\string\citation{#2}}\fi
  \@tempcnta\z@\@tempcntb\m@ne\def\@citea{}\@cite{\@for\@citeb:=#2\do
    {\@ifundefined
       {b@\@citeb}{\@citeo\@tempcntb\m@ne\@citea\def\@citea{,}{\bf ?}\@warning
       {Citation `\@citeb' on page \thepage \space undefined}}%
    {\setbox\z@\hbox{\global\@tempcntc0\csname b@\@citeb\endcsname\relax}%
     \ifnum\@tempcntc=\z@ \@citeo\@tempcntb\m@ne
       \@citea\def\@citea{,}\hbox{\csname b@\@citeb\endcsname}%
     \else
      \advance\@tempcntb\@ne
      \ifnum\@tempcntb=\@tempcntc
      \else\advance\@tempcntb\m@ne\@citeo
      \@tempcnta\@tempcntc\@tempcntb\@tempcntc\fi\fi}}\@citeo}{#1}}
\def\@citeo{\ifnum\@tempcnta>\@tempcntb\else\@citea\def\@citea{,}%
  \ifnum\@tempcnta=\@tempcntb\the\@tempcnta\else
  {\advance\@tempcnta\@ne\ifnum\@tempcnta=\@tempcntb \else\def\@citea{--}\fi
    \advance\@tempcnta\m@ne\the\@tempcnta\@citea\the\@tempcntb}\fi\fi}
\makeatother

\begin{document}
\begin{titlepage}
\pubblock

\vfill
\Title{Physics Case for the International Linear Collider}
\vfill
\Author{LCC Physics Working Group}
\bigskip
\Author{Keisuke Fujii$^1$, Christophe
Grojean$^{2,3}$ Michael E. Peskin$^4$(conveners); Tim
Barklow$^4$, Yuanning Gao$^5$,
Shinya Kanemura$^6$, Hyungdo Kim$^7$, Jenny List$^2$,
Mihoko Nojiri$^1$, Maxim
Perelstein$^8$, Roman P\"oschl$^{9}$,  J\"urgen Reuter$^2$, Frank
Simon$^{10}$, 
Tomohiko Tanabe$^{11}$, Jaehoon Yu$^{12}$
James D. Wells$^{13}$; Hitoshi Murayama$^{14,15,16}$, Hitoshi Yamamoto$^{17}$}

\vfill
\begin{Abstract}
We summarize the physics case for the International Linear Collider (ILC).
We review the  key motivations for the ILC presented in the literature,
updating the projected
measurement uncertainties for  the ILC experiments in accord with the 
expected schedule of operation of the accelerator and the results of
the most recent
simulation studies.
\end{Abstract}

\vfill

\vfill

\newpage
\mbox{\quad}
\vfill

\begin{raggedright}
\noindent $^1$ \kek \\
$^2$  \DESY \\
$^3$  \ifae \\
$^4$ \SLAC\\
$^5$ \Tsinghua \\
$^6$  \Toyama\\
$^7$ \Seoul \\
$^8$ \Cornell\\
$^{9}$  \Orsay\\
$^{10}$ \MPP \\
$^{11}$  \Tokyo\\
$^{12}$ \UTA\\
$^{13}$  \Michigan \\
$^{14}$ \Berkeley \\
$^{15}$ \LBL  \\
$^{16}$ \IPMU\\
$^{17}$  \Tohoku \\
\end{raggedright} 

\vfill

\newpage
\mbox{\quad}

\vfill
\tableofcontents

\vfill

\newpage

\def\thefootnote{\fnsymbol{footnote}}
\setcounter{footnote}{0}
\newpage
\mbox{\quad}

\end{titlepage}

\section{Introduction}

We are now looking forward to the establishment of an international 
collaboration to construct the International Linear Collider (ILC)
in Japan.   The physics potential of this machine is known to be very impressive.
Its capabilities have been
documented,  most recently,  in 
Volume~2 of the ILC Technical Design Report~\cite{ILCTDR}, in a
series of reports to the American Physical Society's study of the
future of US particle physics (Snowmass
2013)~\cite{SnowmassHiggs,Snowmasstop,SnowmassBSM,SnowmassEW}, and in
a comprehensive review article~\cite{Gudi}. 
This article gives  a brief and accessible review of
the main points of these documents. 

The ILC is, at this time, the only energy frontier accelerator for the
post-LHC era that has moved to the engineering stage and has 
 attracted strong attention from a potential host government.
Now, as the project moves to a definite site-specific design,  the
ILC Parameters Joint Working Group
 of the Linear Collider Collaboration has presented a
realistic plan for the operation of the accelerator, including
startup, energy stages, and luminosity upgrades~\cite{ParameterGroup}.
It is appropriate to update the projected capabilities of the
ILC
in accordance with this plan.

In this paper we present a summary of the ILC physics case and
updates of the expected measurement uncertainties, including results from
new simulation studies.   In the program from the ILC Parameters Joint 
Working Group~\cite{ParameterGroup},   the ILC would have an initial phase
in which it accumulates
500~fb$^{-1}$ at 500~GeV, then 200~fb$^{-1}$ at 350~GeV, then
500~fb$^{-1}$ at 250~GeV.  After a luminosity upgrade, the ILC would
accumulate an additional 3500~fb$^{-1}$ at 500~GeV, then an additional
1500~fb$^{-1}$ at 250~GeV.
Results quoted here for the ``initial'' and ``full'' ILC data sets  are
based on this scheme.

The most important aspects of the ILC physics program are: (1)
measurement of the properties of the newly-discovered Higgs boson with
very high precision;  (2) measurement of the properties of the top
quark with very high precision; (3) searches for and studies of new
particles expected in models of physics at the TeV energy scale.  The
specific capabilities of the ILC in these areas are reviewed in the various sections
of this report.   The physics program of the ILC is still broader,
encompassing precision electroweak measurements, 
detailed studies of the $W$ and $Z$ boson couplings, 
tests of Quantum Chromodynamics, and other topics.  A complete survey is
given in Ref.~\cite{ILCTDR}. 

Before we begin, we should make two general points about the role of
the ILC in the current situation in particle physics.  The first is
that the discovery of the Higgs boson at the CERN Large Hadron
Collider~\cite{ATLASdiscovery,CMSdiscovery} is a milestone in the
history of particle physics that changed our perspective on the goals
of this field.  We now have in hand the complete particle spectrum of a
``Standard Model'' that could be correct up to very high energies.
It is possible that this theory of particle physics could be correct
up to energies thirteen orders of magnitude higher than our current
experiments.   However, this would be unfortunate, because this model
is inadequate in several important respects.
First, it does not explain the most basic fact about the Higgs field,
why it is that this field  forms a condensate that fills space and gives rise to
the masses of all known elementary particles.   Second, it has no place for the
particle or particles that make up cosmic dark matter, a neutral,
weakly interacting substance that, according to astrophysical
observation, makes up 85\% of the mass in the universe and 25\%  of its total energy.  Third, it
does not explain the asymmetry in the amount of matter and antimatter
in the universe.  One might add to this list many more fundamental
questions, for example, why quarks and leptons, which make up observed
matter, have the quantum numbers that they do. However, these three questions
are the keys to progress through experiment.  The most pressing
issue in particle physics today is that of where and how the
Standard Model breaks down.  If the questions just listed have answers given
by current theoretical proposals, new particles and forces beyond the
Standard Model should appear at the leading accelerators currently
operated and planned---the LHC and the ILC. 

In the discussion to follow, we will compare the capabilities of the
LHC and the ILC.  However, it is also important to realize that the
experimental programs at these accelerators differ in essential ways.
The LHC gives access to high energies for direct production of new
particles.
However, this comes at a price.  The rates of production of proposed
new particles are typically $10^{-10} - 10^{-12}$ of the
proton-proton total cross section.  Even after selection of
characteristic event types, these processes typically represent only
about 10\% of the total yield, over a background consisting of complex
Standard Model reactions.  This limits both the range of new processes
that can be observed and the precision with which rates can be
measured.  

At the ILC, and more generally in electron-positron
collisions,
 the situation is qualitatively different.  The processes that we wish
to study are large fractions of the total electron-positron
annihilation cross section.  Event selections give high purity, over
backgrounds that are straightforward to compute.  For the study of a 
heavy particle, all decay modes can be observed, and systematic errors on
measured rates are at the 0.1\% level.    This is a powerful and unique
capability that we can apply to the Higgs boson and top quark---the
two known particles most directly connected to the questions
we have listed above---and to any new particles that might appear in
the energy range that the ILC will study. Precision measurements at
the ILC can not only
prove
the existence of new particles with masses well above the $\ee$ collision
energy but also can give detailed information about their properties.
We will see examples of this in all three sections below.

The second point is a perspective on the longer-range future of high-energy
physics.   Our field's need for larger and more powerful accelerators
has driven us to be more globalized than any other field of science.
Today, there is one high-energy proton-proton collider in the world,
the LHC.
Its construction was made possible by the existing complex of tunnels
and infrastructure at CERN.
At the moment, a large fraction of the experimental particle physicists in
the world are collaborators in the two large experiments ATLAS and CMS
at the LHC.   This insures CERN's current stature as the major international
center of particle physics.  

For electron-positron collisions, any
facility at energies much higher than those already realized must be a
linear collider in a long, straight tunnel.   The ILC infrastructure
will provide a basis for collisions at 500\,GeV.   The technology for
ILC can extend the reach of the machine to 1~TeV, as envisioned in the
ILC TDR~\cite{ILCmachine}.   Beyond this, the ILC laboratory will provide a setting
in which 
new generations of technology will provide  electron-positron collisions
at even higher energies.   It will provide a new  world
  laboratory that will be the global host for experiments with electron and
positron beams into the long-term future. 

With this background, we now review the capabilities promised by the
ILC for experiments on the Higgs boson (Section 2), the top quark
(Section 3), and proposed new particles (Section 4).   An appendix
  gives a table of the projected measurement errors for the most
  important parameters.  We recommend that these numbers be used in 
discussions of the ILC physics prospects and in comparisons of the ILC with other proposed facilities.

\section{Higgs Boson}

\subsection{Introduction}

\begin{figure}
\begin{center}
\begin{minipage}[c]{0.50\textwidth}
\includegraphics[width=0.99\textwidth]{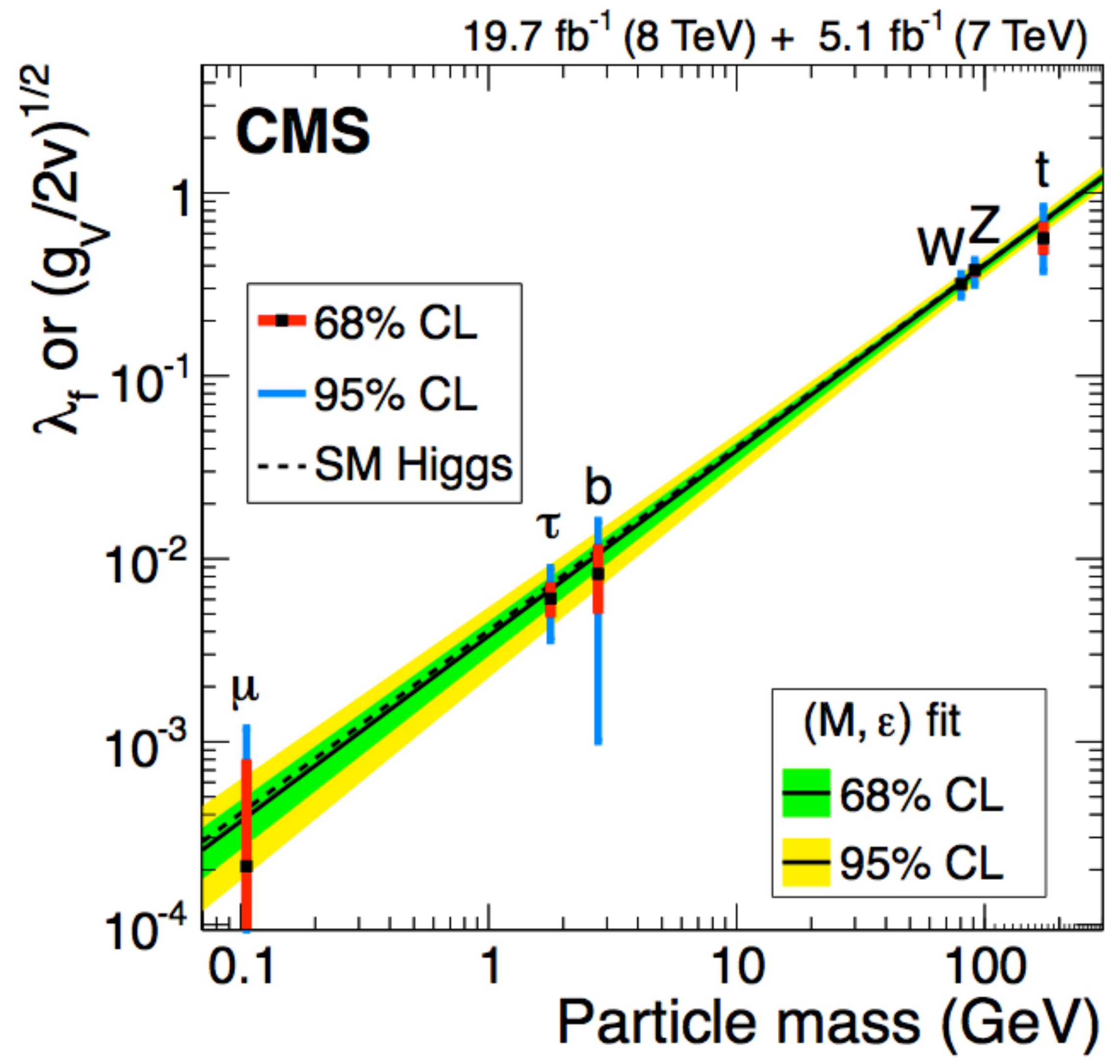}
\end{minipage}
\hspace{0.01\textwidth}
\begin{minipage}[c]{0.47\textwidth}
\includegraphics[width=0.99\textwidth]{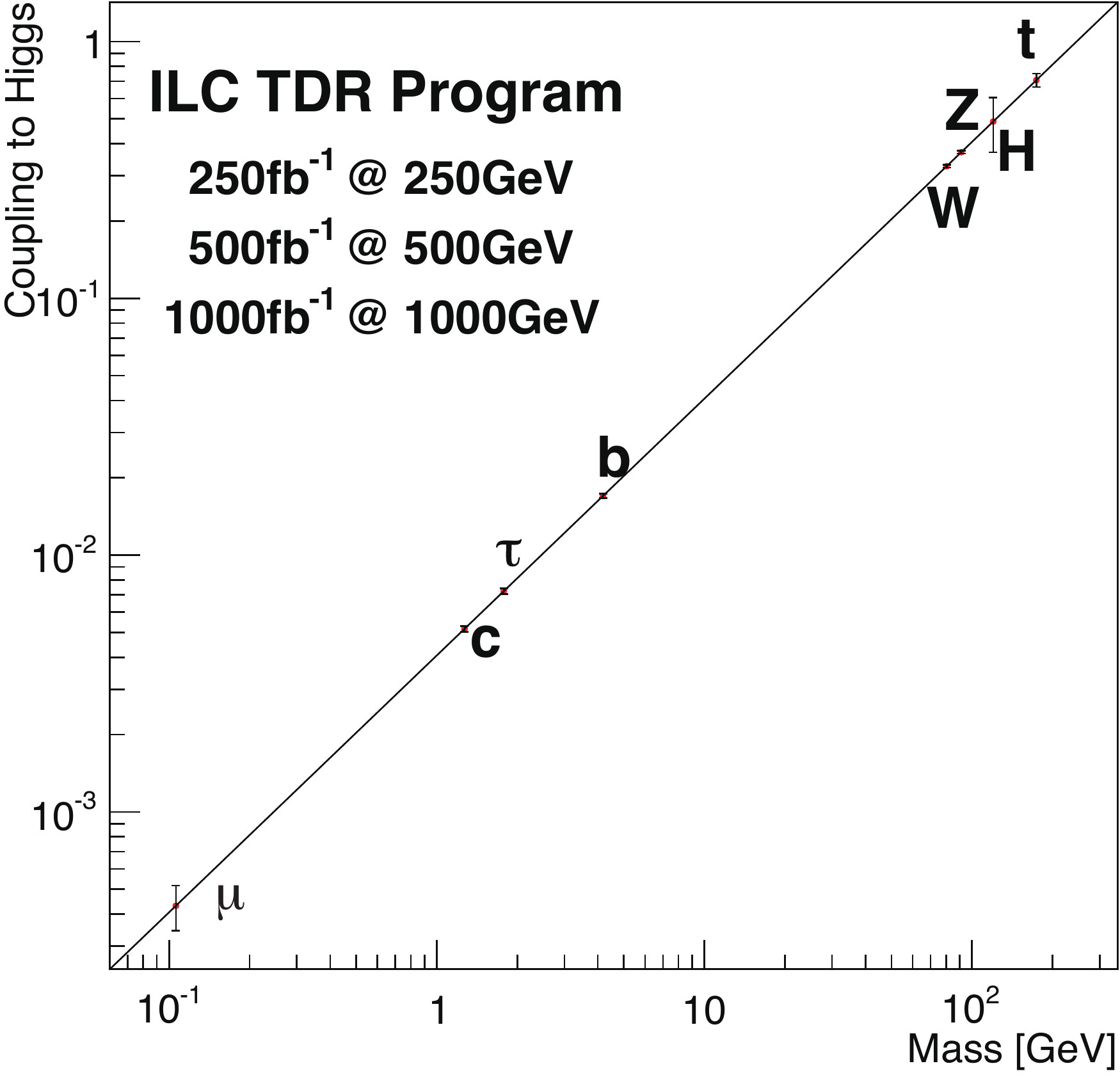}
\end{minipage}
\end{center}
\caption{The Standard Model predicts that the Higgs couplings to
    fundamental fermions are linearly 
proportional to the fermion masses, whereas the couplings to bosons
are proportional to the square 
of the boson masses. Left:  the CMS fit to the current
Higgs data, showing 
consistency with this prediction, from Ref.~\cite{CMSHiggs}. Right: the
expected improvement in the precision in the measurement of 
the Higgs couplings at the ILC, from Ref.~\cite{ILCTDR}.}
  \label{fig:HiggsReggePlot}
\end{figure}

It is a property of the Standard Model of particle physics that,
$10^{-10}$ seconds after the Big Bang, empty space-time made a
transition to a new state filled with a uniform field.  This field,
the Higgs field, breaks symmetries of the Standard Model that forbid
the masses of quarks, leptons, and vector bosons.  Its uniform field value is
therefore responsible for the masses of these particles.  The
observation of the Higgs boson at the LHC gives evidence that this set
of ideas is correct.  In particular, 
the measurements of $Z$ boson
polarization in $h\to ZZ^*\to 4$~lepton events already gives strong evidence that the Higgs boson
couples to the $Z$ boson in the specific manner required to give the $Z$ boson
its mass.  More generally, the LHC measurements of the various
production and decay reactions of the Higgs boson are all consistent
with the picture of the Higgs field as the origin of all elementary
particle masses.  The
current situation is 
illustrated in Fig.~\ref{fig:HiggsReggePlot}(a)~\cite{CMSHiggs}. Over the next decade, it is
likely that the LHC will provide additional 
evidence toward this conclusion.  However, 
this evidence only addresses the question of {\it how} the weak
 interaction symmetry is broken.  It does not address the question of
{\it  why}
 the symmetry is broken or why the Higgs field acquires its nonzero value.

With the measurement of the mass of the Higgs boson, now known to
0.2\% accuracy, the parameters of the Standard Model are 
fixed.  All further properties of the Higgs boson are predicted by the model.
The observation of any deviation from these  predictions  would 
be a clear indication of new physics beyond the Standard Model. 

Different models of new physics beyond the Standard Model lead to 
different patterns of deviation from the predicted Higgs boson couplings.
In supersymmetric models, and more generally in models with more than
one fundamental Higgs field, 
the largest 
deviations are expected to be found in the couplings to the down-type
 quarks and leptons and in the effective couplings to photons and
 gluons.
In models in which the Higgs boson is composite, the effects of
compositeness produce a uniform decrease in all of the Higgs
couplings.
Such models may also have partial top quark compositeness and heavy top
partners; these effects induce further shifts of the Higgs couplings
to top quarks and to photons and
gluons.
Thus, the measurement of the couplings of the Higgs boson will give
evidence on the question of whether the Higgs boson is a fundamental
scalar particle---the first ever observed---or a composite of more
fundamental constituents.
Looking for deviations of the Higgs couplings is also a way to probe
the 
naturalness of the weak scale.
Indeed, general arguments~\cite{Low:2009di,Perelstein} imply that any new physics that 
screens the Higgs mass from large quantum corrections generically
leads
 to deviations in the Higgs couplings to photons and gluons at least as
 large as 1\%.  Supersymmetric and composite Higgs models are prime examples of this general pattern.

However, the size of deviations in the Higgs couplings is limited by
LHC exclusions of new particles and by precision weak interaction measurements.
Taking these constraints into account, the deviations predicted in all of the models above are generically small, at
the level of about 5\% or less, varying as $m_h^2/M^2$, where $M$ is the mass
of the new particles predicted in the model.   The loop-induced
couplings of the Higgs boson to $\gamma\gamma$ and $gg$  receive
contributions from the Standard Model
particles $t$ and $W$, but also, possibly, from new heavy particles.
  In the fits that we present below, we 
consider these couplings to be independent of the couplings of $t$ and
$W$ to the Higgs boson.

At the LHC, the
uncertainties in the Standard Model predictions for the rates of Higgs
processes are also of the order of 5\%, and systematic errors on detection
probabilities are of the same order.   In addition, only a subset of
the Higgs decays can be observed directly.  Because not all Higgs
decays are observed, there are further ambiguities, discussed below. 
 Thus, the goal for Higgs
boson experiments, the measurement of the individual Higgs couplings
to accuracies of better than 1\%, can be met only by experiments at an 
electron-positron collider.  The improvement expected from the ILC over the current
measurements is shown in
Fig.~\ref{fig:HiggsReggePlot}(b)~\cite{ILCTDR}.

\subsection{Higgs boson observation}


As we have discussed above, the ILC will study the Higgs boson using 
the  features available at an $\ee$ collider:  a well-defined initial
state, 
absence of strong-interaction
backgrounds, and controlled and calculable backgrounds from
electroweak processes.   The relatively quiet environment of $\ee$
collisions
also allows the construction of detectors with higher intrinsic
precision and heavy-flavor tagging efficiency  than is possible at the LHC.
These detectors essentially reconstruct
all 
events in terms of fundamental particles such as leptons, quarks, 
and gauge bosons. There are three major Higgs boson production processes
at the ILC: 
$e^+e^- \to Zh$ (``higgsstrahlung''), $e^+e^- \to \nu_e \bar{\nu}_e
h$ (``$W$ fusion''), and  $e^+e^- \to e^+e^- 
h$ (``$Z$ fusion'').  For each of these, we will be able to 
separately identify all of the major Higgs 
decay modes, such as $h \to b\bar{b}$,  $WW^*$, $c\bar{c}$,
$\tau\tau$, 
and $gg$, with high efficiency.
  It is worth recalling that the decays of the
Higgs boson to quarks are very difficult to observe at the LHC.  The
decay $h\to b\bar b$ can be observed only in special kinematics, and
it seems extremely challenging to observe 
 $h\to c\bar c$  or $h\to g g$  (though
the latter coupling can be probed in Higgs production).  On the
  other hand, if one anticipates a special role for the top quark in
  electroweak symmetry breaking, it would be important to measure the
  Higgs coupling to $c\bar c$ as a reference value to understand any
  deviations from the Standard Model predictions in the Higgs
  couplings to $t\bar t$ and $gg$.

\begin{figure}
\begin{center}
\includegraphics[width=0.6\hsize]{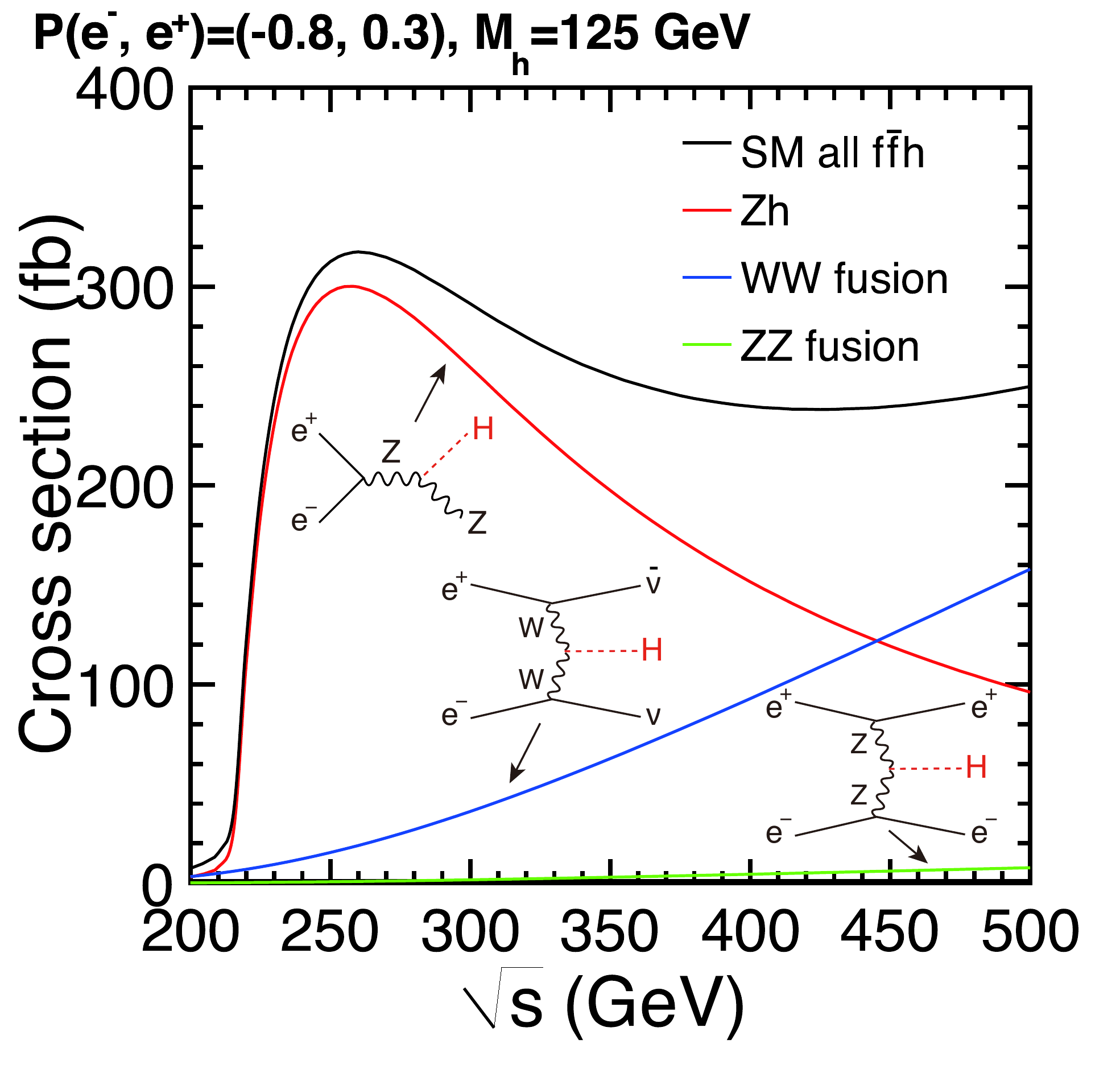} 
\end{center}
\caption{Cross sections for the three major Higgs production processes
 as a function of center of mass energy, from Ref.~\cite{ILCTDR}.} 
\label{fig:sigma_h}
\end{figure}

The control of electron and positron beam polarization that the ILC will make
available allows experimenters to select Higgs reactions or to change
the mixture of signal and background.  For example, 
the $W$ fusion process   $e^+e^- \to \nu_e \bar{\nu}_e h$ 
proceeds only via collisions of left-handed electrons with right-handed
positrons. 
As a consequence, its cross section can be enhanced by a factor of
about 2 with the polarized 
electron and positron beams available at the ILC.
Figure~\ref{fig:sigma_h} plots the cross sections for the single
Higgs boson production 
at the ILC with the left-handed polarization combinations: $P(e^-,e^+)=(-0.8,+0.3)$.
The figure tells us that at a center of mass energy of 250~GeV the
higgsstrahlung process attains
 its maximum cross section, providing about 160,000 Higgs events for
an
 integrated luminosity of $500\,$fb$^{-1}$.  At
500~GeV, a sample of 500~fb$^{-1}$ gives another 125,000 Higgs events,
  of which 60\% are from the $W$ fusion process~\cite{Han}. 
With these samples of Higgs events, we can measure the rates for Higgs
production and decay for all of the major Higgs decay modes.

The higgstrahlung process $e^+e^- \to Zh$  offers another special advantage.   By
identifying the $Z$ boson at a well-defined laboratory energy
corresponding to the kinematics of recoil against the 125\,GeV Higgs
boson, it is possible to identify a Higgs event
{\it without looking at the Higgs decay at all}.    This has three
important 
consequences.   First, as we will describe below, it gives us a way to
determine the total width of the Higgs boson and the absolute
normalization of the Higgs couplings.   Second, it allows us to
observe Higgs decays to invisible or exotic modes.   Decays of the
Higgs boson to dark matter, or to other long-lived particles that do not
couple to the Standard Model interactions, can be detected down to
branching ratios below 1\%.   

\begin{figure}
\begin{center}
\includegraphics[width=0.6\hsize]{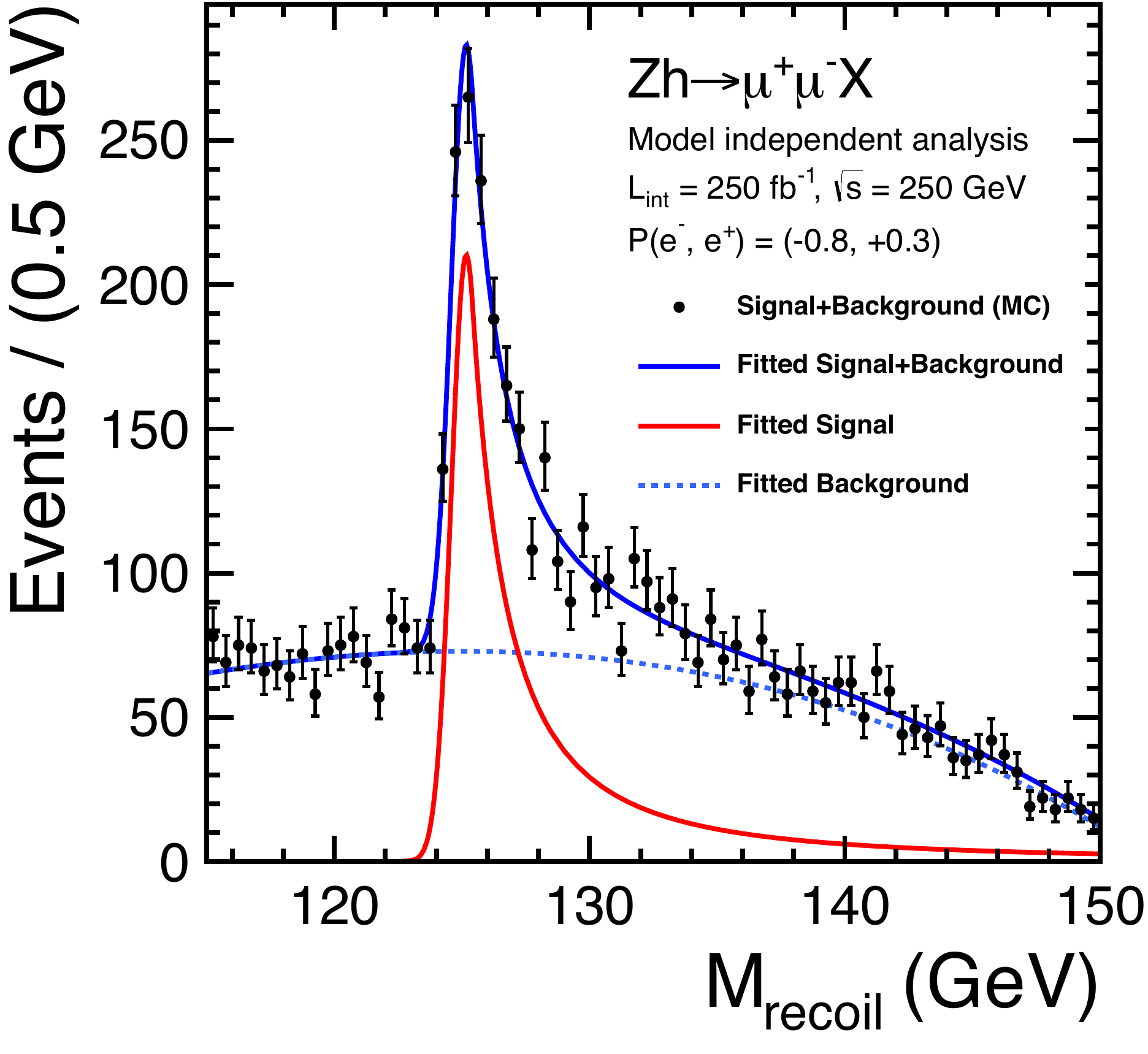}
\end{center}
\caption{Recoil mass distribution for the process: $e^+e^- \to Zh$ followed by $Z \to \mu^+\mu^-$ decay
for $m_h=125\,$GeV with $250\,$fb$^{-1}$ at $\sqrt{s}=250\,$GeV, based
on Ref.~\cite{Li}.} 
\label{fig:mrec}
\end{figure}

Finally, measurement of the decay of the $Z$ to $\ee$ or $\mu^+\mu^-$
gives a very precise determination of the mass of the Higgs boson.
The  mass of a particle recoiling against a lepton pair is given by 
\beq
M_X^2 = \left(p_{CM} - (p_{\ell^+} + p_{\ell^-})\right)^2 ,
\eeq{MX}
where $p_{CM}$ is the 4-momentum of the annihilating electron-positron system.
The expected recoil mass distribution for a 
$m_h=125$\,GeV Higgs boson with 
250\,fb$^{-1}$ at $\sqrt{s}=250$\,GeV is shown in Fig.~\ref{fig:mrec}.
This measurement allows us to determine the Higgs mass to better than
 $30$\,MeV and the cross section to a sub-\% level~\cite{SnowmassHiggs}.
The precision of the cross section can be further improved by adding 
events with decay of the  $Z$ to quarks.

\subsection{Higgs boson coupling measurement}

To compare Higgs boson rate measurements to the Standard Model expectations,
it is important to note that what is actually measured is the rate for
a production and decay process.  This is proportional to the cross
section for Higgs production multiplied by the branching ratio (BR),
which is related to the partial width into the observed channel 
 through the familiar formula
\beq
   BR(h\to A\bar A) =  \Gamma(h\to A\bar A)/\Gamma_h \ , 
\eeq{total}
where $\Gamma_h$ is the total rate of Higgs decay or the total width
of the Higgs boson as a resonance.  In the Standard Model, $\Gamma_h$
is very small, too small to be measured directly as a resonance width.
Unfortunately, we must determine $\Gamma_h$ to learn the absolute sizes of
the Higgs boson couplings.

At the LHC, all determinations of $\Gamma_h$ to high accuracy require model-dependent
assumptions.
At the ILC, however, we can use the fact that all Higgs decay modes
can be identified in  the higgsstrahlung process using the recoiling
$Z$ boson to measure certain Higgs
couplings directly.    The total rate for higgsstralung is
proportional to the $ZZh $ coupling.  The rate for the $W$ fusion process
\beq
      \ee \to \nu\bar\nu h \ \quad \mbox{with} \quad   h\to b\bar b\ ,
\eeq{WWbbrate}
divided by $BR(h\to b\bar b)$ determined with higgsstrahlung, gives
the $WWh$ coupling.   These measurements then determine $\Gamma(h\to ZZ)$
and $\Gamma(h\to WW)$.   Combining these results with the measured
branching ratios using \leqn{total}, the ILC measurements give a
model-independent determination of $\Gamma_h$.  That result in turn
fixes the absolute size of all other Higgs couplings.

The most statistically powerful determination of the Higgs width
$\Gamma_h$ uses the $W$ fusion process, which turns on at 
$\ee$ center of mass energies above 250\,GeV, as  shown in Fig.~\ref{fig:sigma_h}.
Thus,  the most precise coupling measurements from the ILC
require data-taking at energies of 350\,GeV or above.
 The coupling precisions can be further improved by increasing the
 data sample or by running at still higher energies.  

Because the decay of the Higgs boson to $\gamma\gamma$ is rare, with a
branching ratio of 0.2\%  in the Standard Model, it will be difficult
for the ILC to gather large statistics for this decay.  Fortunately,
the $\gamma\gamma$ and $ZZ$ decay modes of the Higgs boson are the modes
that are most straightforward for the LHC experiments.
The LHC  is expected to measure the ratio of branching ratios 
$BR(h \to \gamma\gamma)/BR(h \to ZZ)$ very accurately, using 
a technique in which the systematic errors largely cancel.    In
Ref.~\cite{ATLAS}, 
it is estimated that this ratio of branching ratios can
be measured to 2\% accuracy.
Combination of this with the $ZZh$ coupling measurement from the ILC
will allow us to reach the required 1\%-level precision 
also for the $h \to \gamma\gamma$ coupling~\cite{MEPHiggs}.  This synergy  is
illustrated 
in the $\gamma\gamma$ entries of  the summary figures cited below.

\begin{figure}
\begin{center}
\includegraphics[width=0.7\hsize]{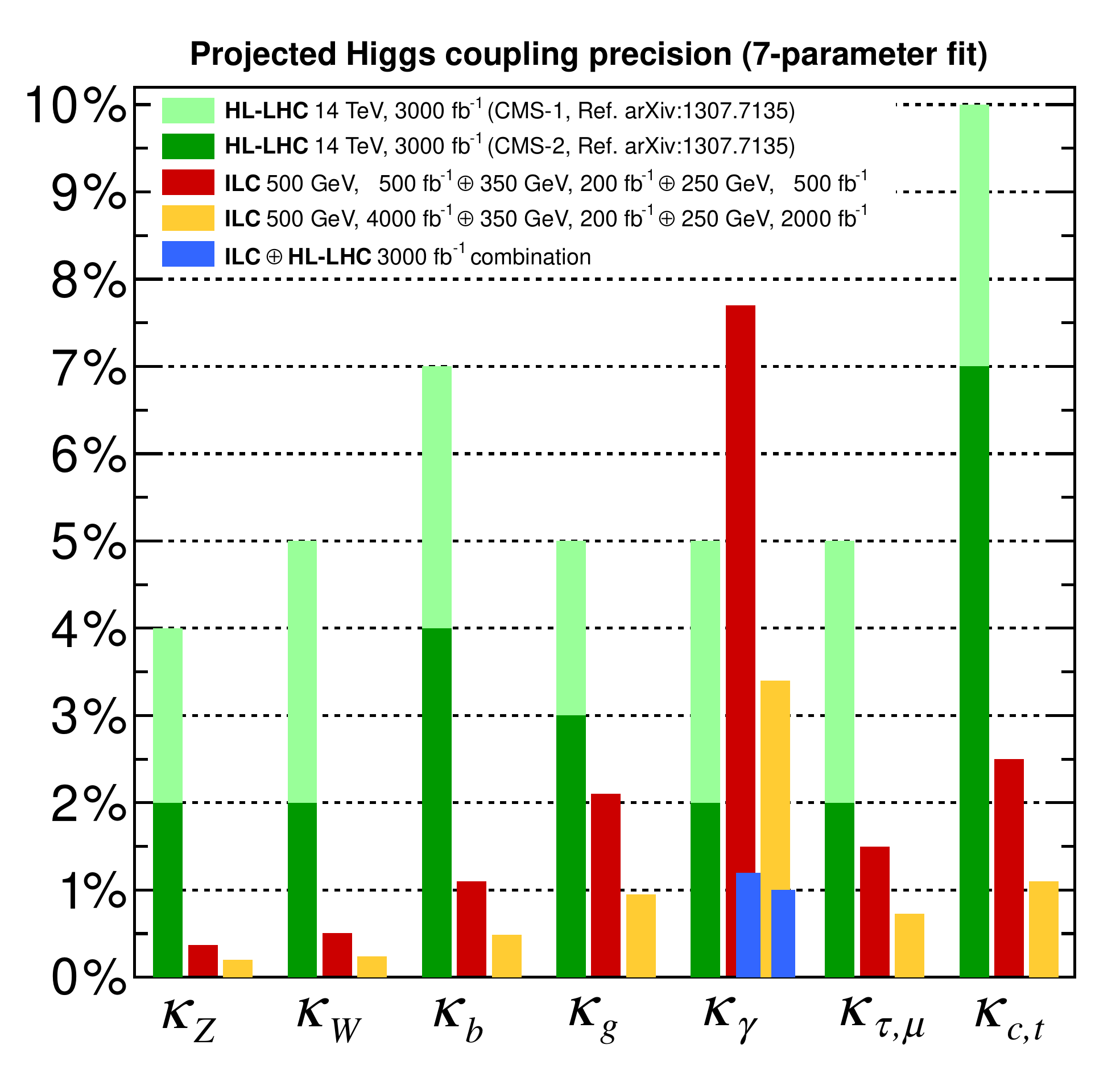}
\end{center}
\caption{Relative precisions for the  various Higgs
  couplings extracted using the model-dependent fit used in 
  the Snowmass 2013 study~\cite{SnowHiggsReport}, applied
 to expected data from the
  High-Luminosity LHC and from the ILC.
Here, $\kappa_A$ is
the ratio of the $A\bar A h$ coupling to the Standard Model
expectation.  The red bands show the expected errors from the initial
phase of ILC running.  The yellow bands show the errors expected from
the full data set.  The blue bands for $\kappa_\gamma$ show the effect
of a 
joint analysis of High-Luminosity LHC and ILC data. }
\label{fig:HiggsILCLHC}
\end{figure}

Since the top quark is the heaviest particle in the Standard Model and
hence most strongly coupled to the Higgs sector,  the Higgs boson
coupling to the top quark could contain special effects and should be
measured independently.
This coupling is 
not directly accessible from the Higgs decay measurements. To determine
it,  we use the reaction $\ee \to t\bar{t}h$ with $h$
observed in its $ b\bar {b}$ decay. The rate of this reaction is
proportional
to $BR(h\to b\bar b)$, but that quantity will have been measured very 
accurately in the program described above.
We can then determine the Higgs coupling to top quarks
 by just counting the number of $t\bar{t}h$
events.  Full-simulation studies for ILC at 500\,GeV show~\cite{ILCTDR}
 that this will provide a 6.3\% measurement of 
the $t\bar t h$ coupling  with the full ILC data set. It is worth pointing out that
the energy of 500~GeV is very close to the production threshold for
this process.
Thus, the full ILC energy of 500~GeV is necessary to achieve these
goals.  On the other hand,
a slight increase of the center of mass energy, by 10\%, 
enhances the cross section by a factor of about four and improves the
precision to 3\% for the same integrated luminosity.

In Fig.~\ref{fig:HiggsILCLHC},  we compare  the uncertainties
in Higgs couplings expected from the High-Luminosity LHC and from the
two phases in 
the evolution of the ILC program.  Because the LHC experiments cannot measure all
Higgs decay modes, they cannot make a model-independent determination
of the Higgs width $\Gamma_h$.  Thus, in  this figure, the 
couplings are determined with constraints that fix the unobserved modes.
Following Ref.~\cite{SnowHiggsReport}, this fit assumes that the
fractional shift in the Higgs couplings is equal for $u,
c, t$, for $d,s,b$, and for $e,\mu,\tau$, and that there is no Higgs
decay to 
invisible or exotic particles.
The large green bars give the
uncertainty projections from the CMS Collaboration assuming current values of systematic
errors.   The heavier green bars assume that systematic errors can be
decreased by the same factor as statistical errors, by a factor of 12
from today to the end of the High-Luminosity LHC program~\cite{CMS}
Projections by the ATLAS collaboration are similar~\cite{ATLAS}.  The ILC estimates
are based on  current full-simulation studies and the ILC program described
in Ref.~\cite{ParameterGroup}.

\begin{figure}
\begin{center}
\includegraphics[width=0.9\hsize]{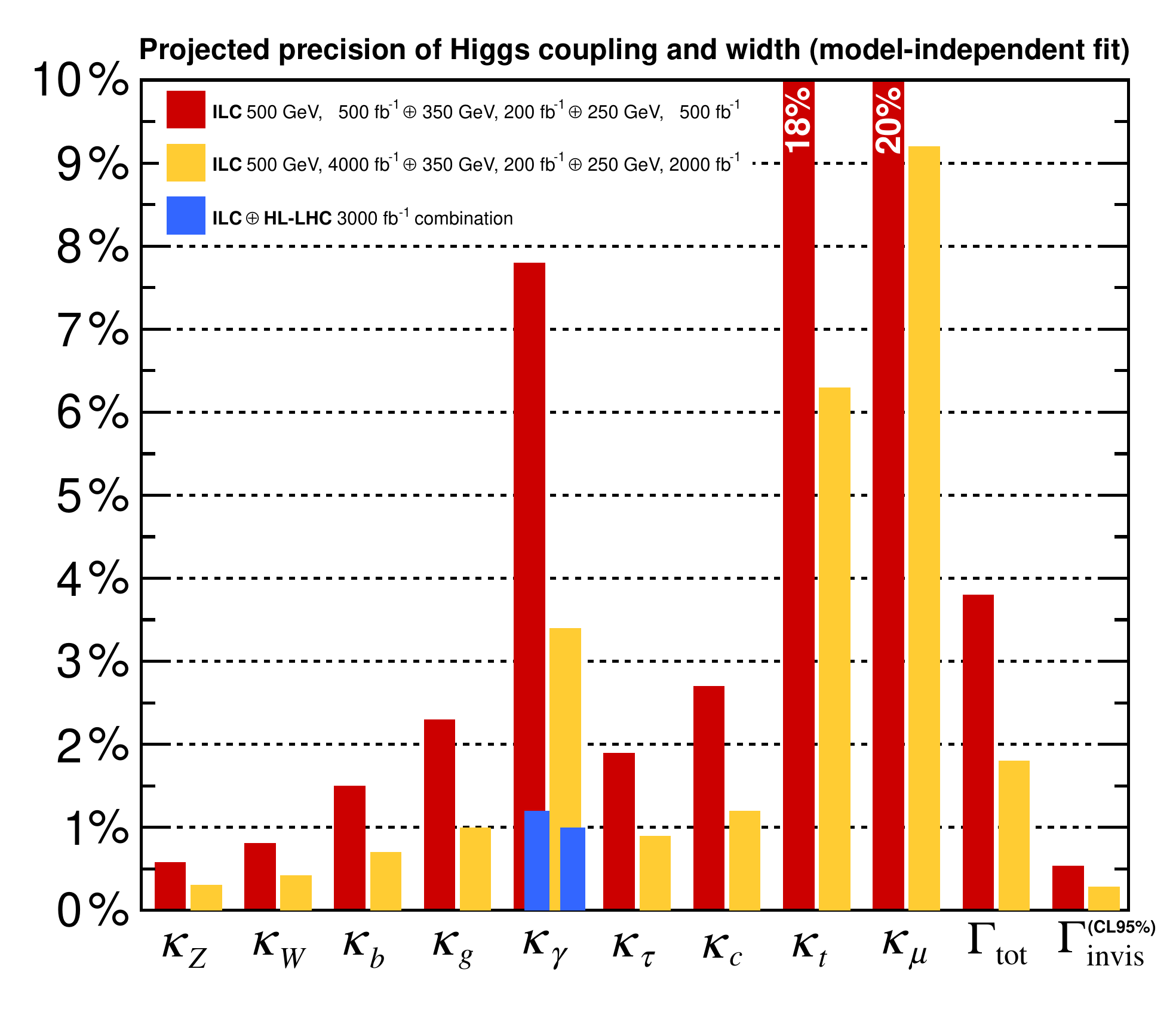}
\end{center}
\caption{Relative precisions for the  various Higgs
  couplings extracted from a model-independent fit to expected data from the
  ILC. The notation is as in Fig.~\ref{fig:HiggsILCLHC}.}
\label{fig:couplings}
\end{figure}

Figure~\ref{fig:couplings} shows the estimated
uncertainties from the ILC for a model-independent fit to the Higgs
boson couplings in which all Higgs couplings, including couplings to
inivisible and exotic modes,  are separately taken as
free parameters.    We see that, in these model-independent
determinations, most couplings 
reach  the
required   precision of 1\% or better in the course of the ILC
program.  As noted above, running the ILC  at 550\,GeV rather than 500\,GeV would
give precisions of 9\% and 3\% in the two entries for the $t\bar
t h$ coupling.

\begin{figure}
\begin{center}
\begin{minipage}[c]{0.48\textwidth}
\includegraphics[width=0.99\textwidth]{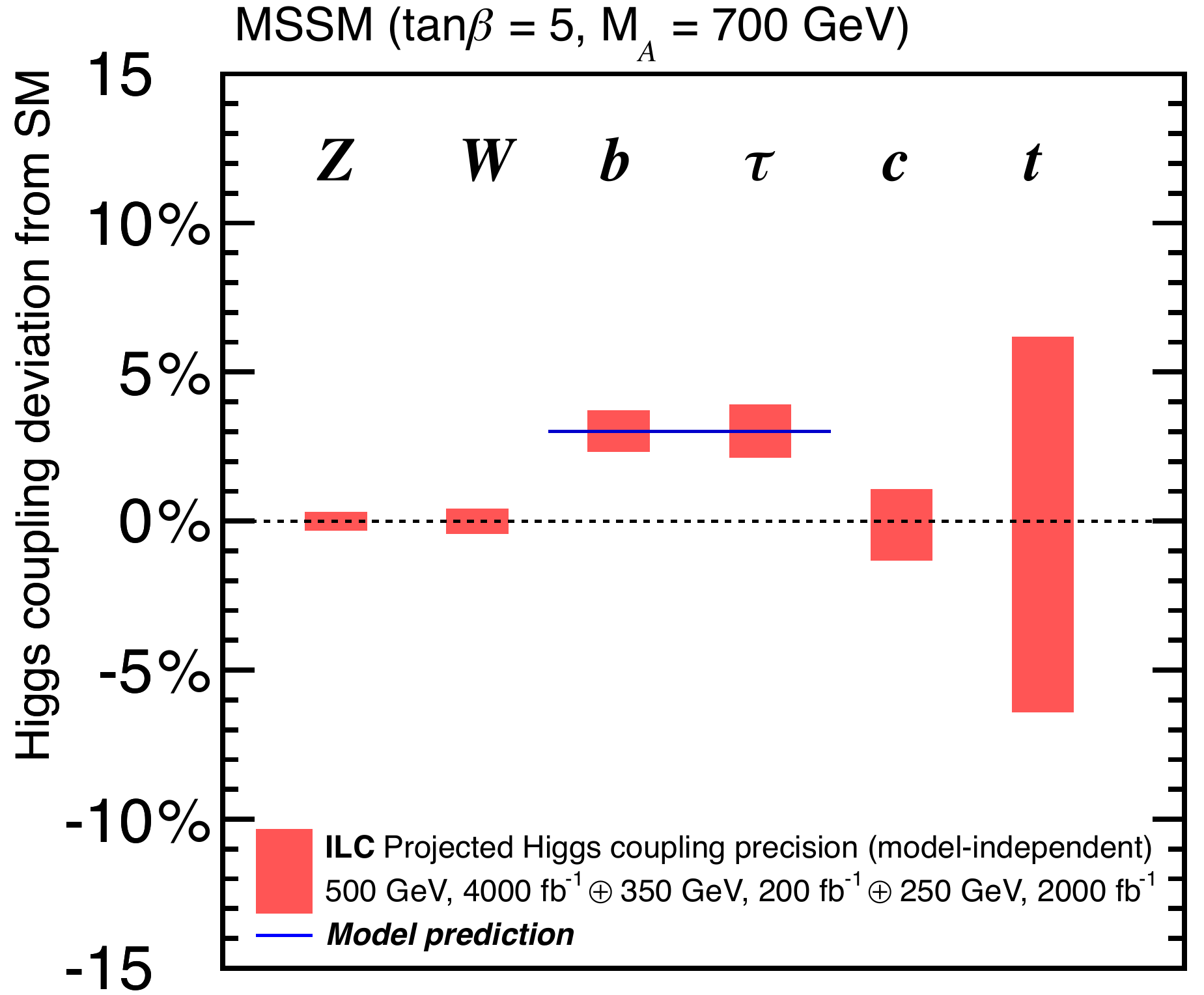}
\end{minipage}
\hspace{0.01\textwidth}
\begin{minipage}[c]{0.48\textwidth}
\includegraphics[width=0.99\textwidth]{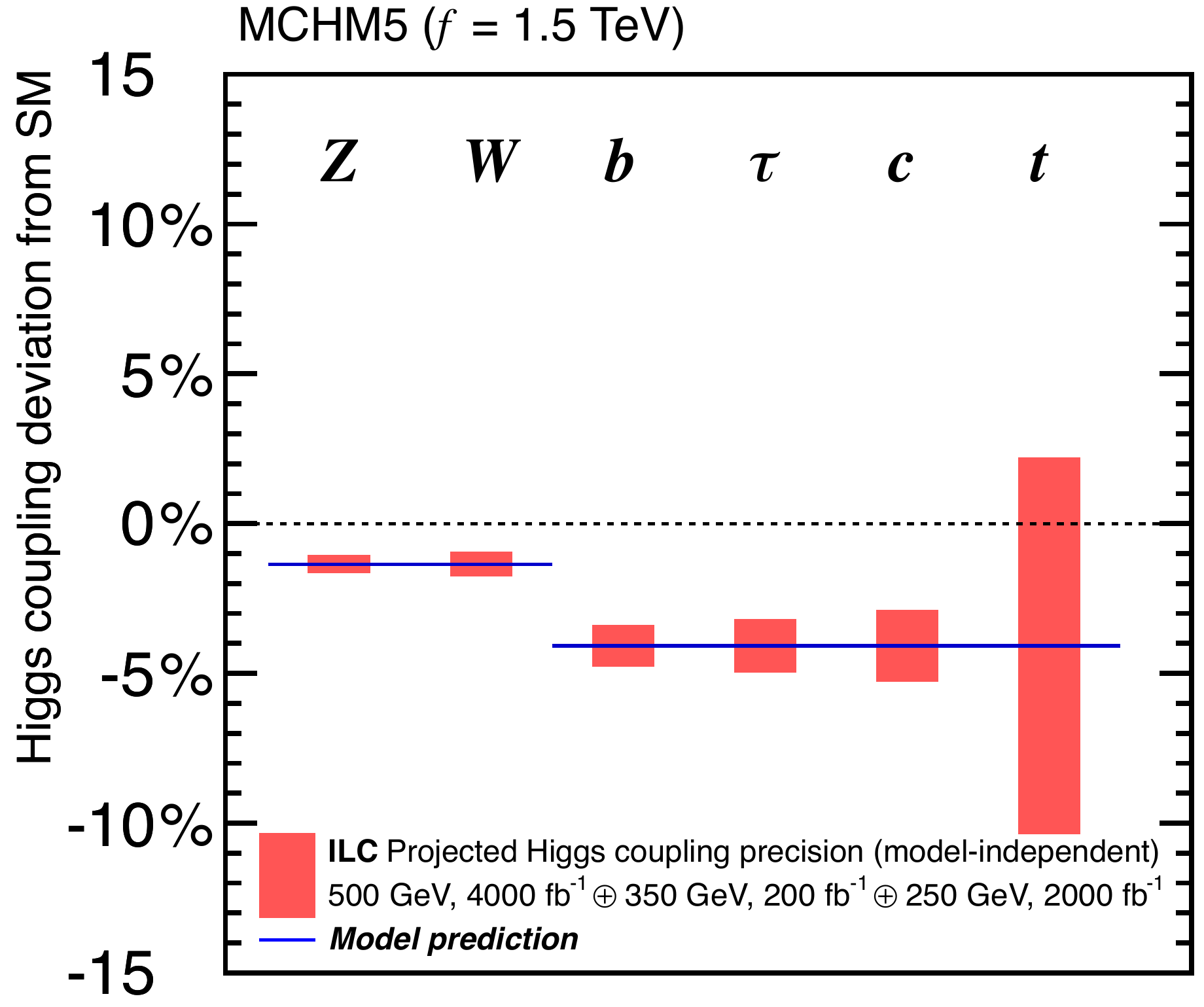}
\end{minipage}
\end{center}
\caption{Two examples of models of new physics and their predicted
  effects on the pattern of Higgs boson couplings.  Left: a supersymmetric model. 
  Right: a model with Higgs boson compositeness.   The 
error bars indicate the 1$\sigma$ uncertainties expected from the
model-independent fit to the full ILC data set.}
\label{fig:fingerprinting}
\end{figure}

Figure~\ref{fig:fingerprinting} shows the power of the ILC precision
to 
distinguish different  models of new physics 
through Higgs boson measurements.  The two panels
illustrate the shifts in the Higgs couplings
from the Standard Model predictions expected in two representative
models of new physics.   The error intervals are those expected from 
the full ILC data set using a model-independent analysis.  Similar illustrations for
additional
models of new physics are presented in Ref.~\cite{ref:kanemura2013}.  It is important not only to
observe deviations from  the Standard Model but also to use the observed
pattern
of deviations as a clue to the correct model that lies behind it.
The ILC will give us that capability.

\subsection{Higgs boson self-coupling}

There is one more important Higgs coupling not shown in
Fig.~\ref{fig:couplings}.
This is the trilinear Higgs self-coupling, which determines the shape
of the potential energy for the Higgs field.   The value of
this coupling gives evidence on the nature of the phase transition in the
early universe  from
the symmetric state of the weak interaction theory to the state of
broken symmetry
with a nonzero value of the Higgs field.

%
In the Standard Model, this transition is predicted to be continuous~\cite{Kajantie:1996mn}. 
However, if the transition were first-order, it would put the universe
out of
 thermal equilibrium and, through possible CP violating interactions
 in the
 Higgs sector, it would allow the generation of a nonzero
 baryon-antibaryon asymmetry.
This is not the only theory for the baryon-antibaryon asymmetry,
 but it is the only theory in which all relevant parameters can 
potentially be measured at accelerators, setting up a quantitative
experimental test. 

The first step would be to test the nature of the phase transition. 
Models in which the phase transition is first-order typically require the Higgs
self-coupling to differ from the value predicted by the Standard
Model~\cite{h3EWPT}.  The Higgs self-coupling can be a factor of 2 larger in some
models~\cite{Aoki}.



 At the High-Luminosity LHC,  double Higgs production can be
detected in well-chosen final states, for example, the state in which one Higgs boson decays to
$\gamma\gamma$, providing a clean signal, while the other decays to
$b\bar b$, providing the maximum rate.   This process should
eventually be observed at the LHC, though current fast-simulation studies are 
rather pessimistic~\cite{ATLASfuture}.  

At the ILC at 500~GeV, pairs of Higgs bosons are produced through
  $\ee \to Z h h $. All Higgs decay modes are observable and will contribute to the
measurement. The modes $hh\to b\bar{b}b\bar{b}$ and $hh\to b\bar{b}WW $ have been
studied in full simulation at the center of mass energy 500\,GeV. Combining the
preliminary results of these two channels only, these  ILC simulations
currently predict a precision of 27\% on
the Higgs self-coupling with the full ILC data set.  This would
already be more than $3\sigma$
evidence for the existence of the Higgs self-coupling at the Standard
Model value. It gives 
a substantial discovery potential for models of the Higgs phase
transition that  predict a larger
value.  Further improvements from the inclusion of more decay modes and refined
analyses are under study.   

Running at higher energy allows one to study the process $\ee
  \to \nu\bar \nu  h h $, whose cross section increases with energy
  and has a different functional dependence on the
  self-coupling from 
  the $Zhh$ reaction.  The decay 
mode $hh\to bbbb$ has been studied in full simulation at
 1~TeV.  Using both the $hh\to bbbb$ and $hh\to bbWW$ modes at
 1~TeV, we expect a precision on the Higgs boson self-coupling of 
16\% for 2000\,fb$^{-1}$ and 10\% for 5000\,fb$^{-1}$~\cite{Tian,Kurata}.

\section{Top Quark}

Among the six quarks of the Standard Model of particle physics, the
top quark has a special role.    It is the heaviest of the six and, we
now know from the LHC, the heaviest particle with the quantum numbers of
a Standard Model quark.   By virtue of its large
mass, the top quark has the strongest coupling of any known particle
to the Higgs field or fields that generate the spontaneous breaking of
the weak interaction symmetry.  

 The top quark has a central position in 
all models of new physics beyond the Standard Model.  Such models must
contain new particles that are, in a well-defined sense, partners of
the top quark.   In the most important models, including supersymmetry
and models with new space-time dimensions, it is the coupling of the
Higgs fields to the top quark and its partners that causes the Higgs
field to develop a symmetry-breaking value in all of space. 

Through particle physics experiments, we have a very detailed
knowledge of  the other heavy quarks, $c$ and $b$.   The information
about these quarks comes from both hadron and from $\ee$ colliders.
In general, hadron colliders supply information on rare processes and
on the quark-gluon coupling, while $\ee$ colliders supply information
on the weak and electromagnetic couplings.  Hadron colliders require
specific, relatively simple, decays to recognize heavy quarks, while
at $\ee$ colliders, heavy quark production provides a large and well
characterized part of the total event rate.    For this reason, 
 the full, very rich, pattern of weak interaction decays of
$c$ and $b$ has been observed mainly at $\ee$ colliders.    Though the
lifetime of the $b$ quark was measured very accurately at hadron
colliders, the pattern of CP-violating couplings of the $b$ quark were first
measured at $\ee$ colliders at KEK-B and PEP-II. 
  These experiments gave the crucial evidence for the
Kobayashi--Maskawa model of CP violation~\cite{Bfact,Bbook}.

For the top quark, we have today only the hadron collider side of the
story.   The pair production threshold of the top quark is around 350
GeV, higher than the center of mass energy of any $\ee$ collider that
has operated so far.    The top quark was discovered at the Fermilab
Tevatron and has been studied with high statistics at the LHC.
However, the difficulties of recognizing and reconstructing top quark
events in the presence of large backgrounds, as well as theoretical
uncertainties associated with
 the interpretation of the hadron collider data, limit the accuracy of
measurements on this particle.    The ILC will allow us, for the first
time, to study the top quark in $\ee$ collisions, where we will be
able to access the widest variety of final states with high fidelity.

There are two somewhat distinct physics programs on the top quark at
the ILC.   The first is the study of the threshold for $t\bar t$
production.  This is the ``hydrogen atom of strong interactions'', the
first situation in which bound states predicted by QCD
  can be studied in a setting
that is free of the nonperturbative, quark-confining, part of the
interaction.     ILC
measurements near 350~GeV will test this theory and also measure the
mass of the top quark to a precision below $10^{-3}$.  In addition,
other properties
 such as the total width will be 
 accessible via
 these measurements. The second is   
the study of top quark production and decay at 500~GeV.  These
measurements will make use of accurate reconstruction of $t\bar t$ events to probe
the full structure of the top quark coupling to the electroweak
interactions and
 provide excellent sensitivity to physics beyond the Standard Model.

\subsection{Top quark at threshold}

\begin{figure}[t]
\begin{center}
\includegraphics[width=0.75\hsize]{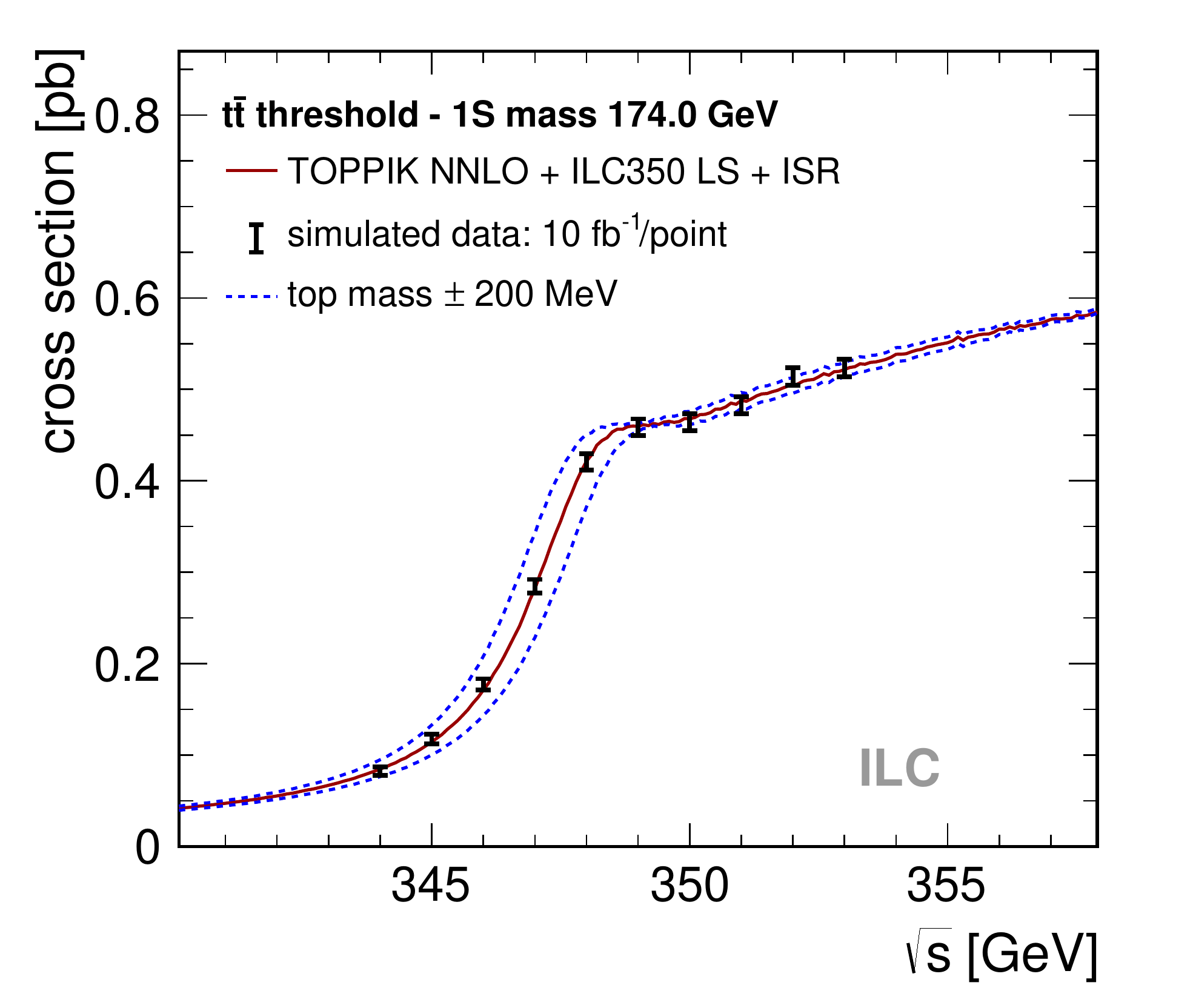}
\end{center}
\caption{Top quark pair production threshold, including the luminosity
 spectrum of the ILC, and simulated data points, corresponding in
 total
 to one year at design luminosity, from Ref.~\cite{Seidel:2013sqa}.}
\label{fig:threshold}
\end{figure}

If the top quark were stable, the $t\bar t$ system would show
prominent resonances at the 1S, 2S, {\it etc.}, quark-antiquark bound
states.   The QCD potential between $t$ and $\bar t$ is expected to be
approximately Coulombic.  The Bohr radius is expected to be
sufficiently small that the nonperturbative, confining part of the
potential would  have a negligible effect on the properties of
the lowest bound states. In reality, the top quark decays, with a
width predicted to be 
about 1.4~GeV. However, this width serves only to smear out the
resonances in a well-defined way without adding new ambiguities.
 The theory of the top quark threshold has been worked out to high
 precision~\cite{ttbarthreshreview,WHIZ,Beneke}, so very accurate predictions are
available to compare to experimental
measurements.  
The predicted threshold shape and simulated ILC measurements are shown
in Fig.~\ref{fig:threshold}.

The top quark threshold region occupies an interval of less than 10\,GeV in the
energy of the $t\bar t$ system.  The study of the shape of the
threshold is completely inaccessible to measurements at hadron
colliders, due to the poor definition of the parton-parton  center of
mass energy in 
hadronic reactions.

 The real part of the pole corresponding to the 1S bound state is a
precisely defined quantity that can be extracted from the threshold
measurements.  This mass parameter can be determined to about 50~MeV
in the ILC program.  The accuracy of this measurement  is limited by
the precision of the
theoretical prediction of the threshold shape, now known at
N$^3$LO~\cite{Beneke,Beneketwo}. 
 For the 200~fb$^{-1}$ data set expected
near 350~GeV~\cite{ParameterGroup}, the expected statistical
errors  in a
3-parameter fit to the threshold shape are    17~MeV for $m_t$,   26~MeV
for $\Gamma_t$, and 4.2\% for the top quark Yukawa
coupling~\cite{Seidel:2013sqa,Horiguchi}. 
 Uncertainties from
knowledge of 
the ILC beam parameters are expected to be still smaller. 

 The 1S top quark mass is  connected to other theoretically precise
definitions of the top quark mass, such as the $\overline{MS}$ mass, to an
accuracy of about 10\,MeV~\cite{Marquard}.    The error in converting
an on-shell top quark mass to the $\overline{MS}$ mass is more than an
order of magnitude greater.  Further, the 
mass usually quoted from Tevatron and LHC data is simply the input
value used in a popular Monte Carlo event generator; its connection to
theoretically precise values is not understood.  At the
High-Luminosity LHC, it is estimated that the $\overline{MS}$ top
quark mass can be extracted to an
accuracy of about 500\,MeV in a theoretically precise way by
measuring the jet-lepton endpoint in leptonic top decays~\cite{topworking}.

The top quark mass is a basic input parameter for the Standard Model.
Other precision tests of the Standard Model are compared to
predictions that require an accurate value of the top quark mass.  For
example, an error of 600\,MeV in the top quark mass corresponds to an
error of 5\,MeV for the prediction of the mass of the $W$
boson.
At the ILC, we expect to measure the mass of the $W$ boson to a few
MeV, a level that gives sensitivity to loop corrections from a variety
of predicted new particles~\cite{EWworking}. 

A precise knowledge of the top quark mass is also relevant to an
unusual
prediction of the Standard  Model.  If there are no new particles
below $10^{16}$\,GeV, the Standard Model predicts that the potential
for the Higgs field turns over at large values and eventually becomes negative.   Then
our universe is unstable over long time scales with respect to tunnelling to a
ground state in which the Higgs field takes an extremely large vacuum
value.   The instability is driven by the interaction between the
Higgs field and the top quark.   The instability occurs if the top
quark mass is greater than 
171.1\,GeV, a value about 2$\sigma$ below the 
currently measured value~\cite{instability}.   So, even if the
Standard Model were literally correct, we would need to improve the
measurement of the top quark to be  confident of the     ultimate fate
of the universe.

\subsection{Top quark weak and electromagnetic couplings}

\begin{figure}
\begin{center}
\includegraphics[width=0.7\textwidth]{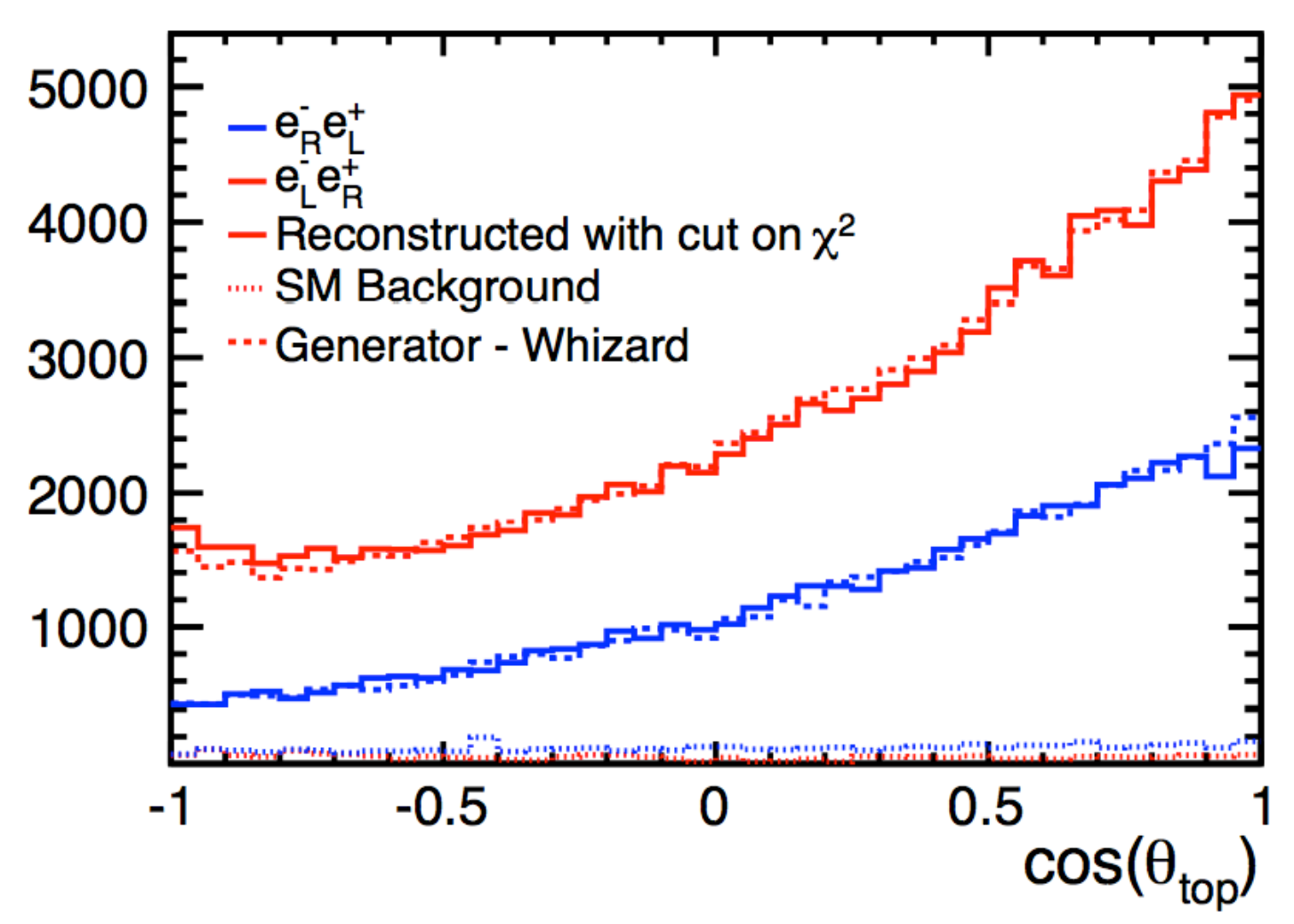}
\end{center}
\caption{The
  angular 
distribution of produced top quarks obtained from full simulations based on a realistic
detector 
model, full event reconstruction, and the inclusion of 
physics and machine-related backgrounds, compared to the 
corresponding generator-level distributions, from Ref.~\cite{Amjad}}
\label{fig:ttpair}
\end{figure}

 At higher energies, top quark and antiquark pairs are produced in the
continuum through $s$-channel $\gamma$ and $Z$.   The
contributions from the $\gamma$ and $Z$ diagrams have ${\cal O}(1)$
interference, which is constructive or
destructive depending on the electron and positron beam polarizations
and the top quark polarizations. This generates ${\cal O}(1)$
forward-backward and polarization asymmetries, shown in Fig.~\ref{fig:ttpair}, and also
parity-violating asymmetries in the top quark decays.  
The top and antitop decays through
weak interactions serve as polarization analyzers.

The central feature of the weak interaction, the feature that
distinguishes it from the strong and electromagnetic interactions, and
the feature that required the intervention of the Higgs field, is that
the couplings depend on polarization. Making use of the unique
capability of the 
ILC for polarized electron and positron beams, we will be able to measure the individual
couplings of each polarization state of  the top quark to the weak
interaction bosons $W$ and $Z$.  The
measurement accuracies from the ILC should improve by about an order of magnitude
over what is projected for the LHC.  The discrimination of the left-
and right-handed couplings to the $Z$ boson is a unique feature of the ILC
measurements.   ILC can also unambiguously determine the signs of the
two couplings.   With the full ILC data set, the experiments should
  achieve a relative precision of 0.6\% on the coupling of the
  left-handed top quark and 1.0\% on the coupling of the right-handed
  top quark~\cite{Devetak,Amjad,Richard,Roman}.

\begin{figure}[t]
\begin{center}
\includegraphics[width=0.8\hsize]{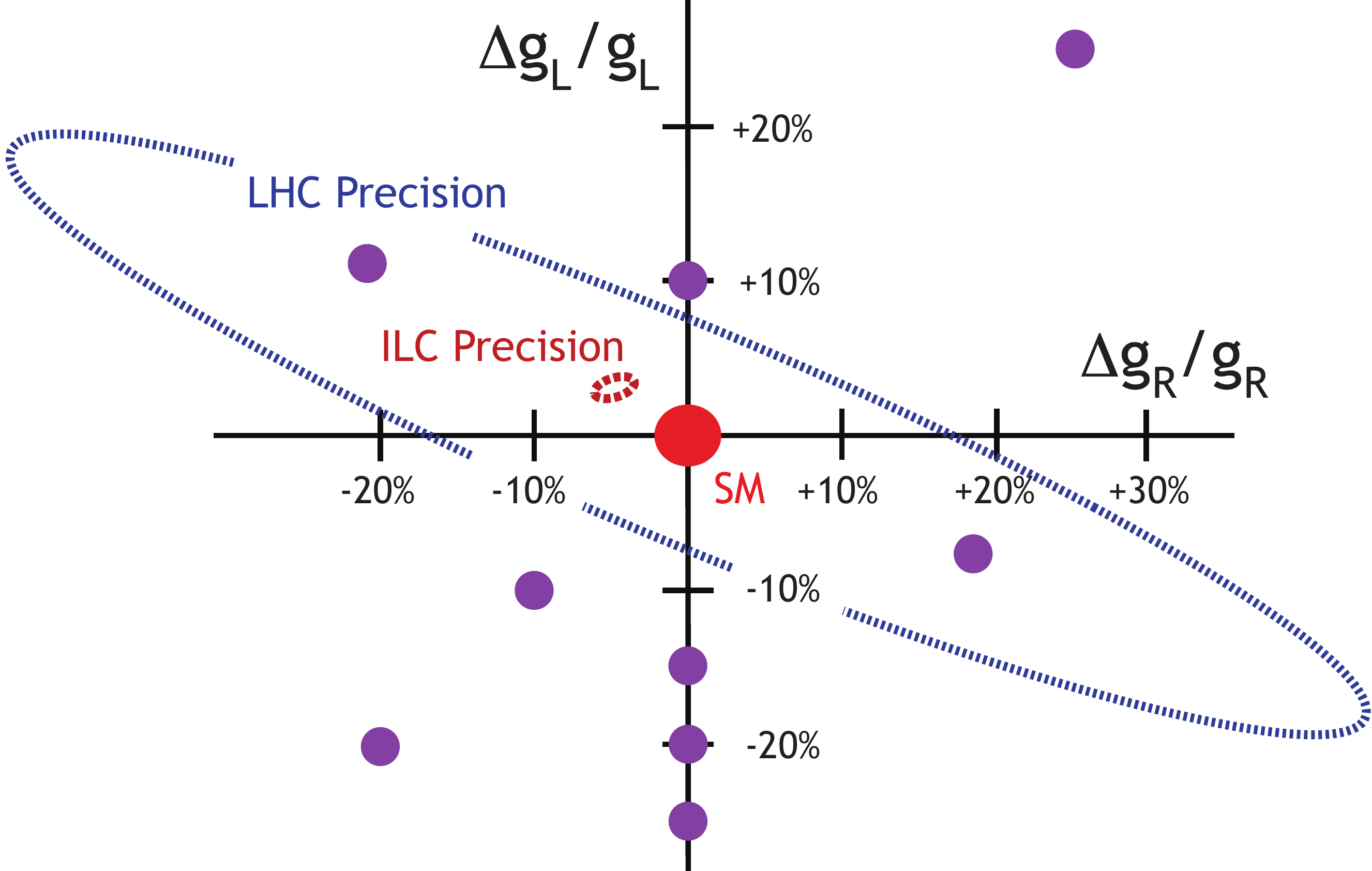}
\end{center}
\caption{The heavy dots display the shifts in  the left- and right-handed
  top quark couplings to the $Z$ boson predicted in a variety of models with
  composite Higgs bosons, from Ref.~\cite{Richard}.  The ellipses show the 68\%
  confidence regions for these couplings expected from the 
  LHC~\cite{topworking,Rontsch} and the  ILC~\cite{Roman}.}
\label{fig:ttcouplings}
\end{figure}

These polarization-dependent couplings receive corrections in most
models of new physics beyond the Standard Model.  The effects are
particularly large in models in which the Higgs boson is a composite
built of some more fundamental constituents.   In such models, the
shifts of the $t\bar t Z $  couplings can be 20\%  or larger and are
expected to be different between the couplings to the two top quark
polarization states.   Figure~\ref{fig:ttcouplings} shows
a survey of theoretical predictions collected in Ref.~\cite{Richard}. 
 The separate values of these couplings provide a
powerful diagnostic of the model.  The measurement 
accuracies expected
at the ILC and the LHC are also shown in the figure.  Measurements
with the ILC accuracy will not only establish the shifts of the $Z$
couplings with high
significance but also pin down properties of the model that gave rise
to them.    A 1\% measurement of these
couplings is sensitive, in models of this type, to the presence of a
10--15~TeV Higgs-sector resonance coupling to $t\bar t$.   This goes 
beyond the ultimate reach of the High-Luminosity LHC for direct
searches for such a resonance,  estimated to be about 5--6 TeV~\cite{topworking}. 

Full reconstruction of top quark pair events at the ILC will also allow
us to search for exotic decay modes of the top quark, and for 
nonzero magnetic and electric dipole moments.
   The latter measurements provide a unique and powerful
probe of CP-violating interactions of the top quark~\cite{TESLA,Hioki}, which provide the
driving force in one
class of models of the cosmic matter-antimatter asymmetry.

\section{New Particles}

In addition to searches for new particles and forces through the
precision study of the Higgs boson and the top quark, the ILC will
carry out direct searches for new particles outside the Standard
Model.
The LHC has already carried out a broad program of searches for new
particles, setting upper limits on masses higher than   1~TeV in
the best cases.  Still, it is possible that  new particles are being 
produced at the LHC and yet are not visible to the experiments there.
Such particles do not appear only in artificial examples but even in some
of the best-motivated scenarios for new physics.  We will review some
specific
models of this type below.   At the ILC, we can
use the advantages of $\ee$ collisions to discover or definitively
exclude these particles.  

A new capability that the ILC will make available is the ability to
polarize the colliding  electron and positron beams.   We have already discussed
the use of beam polarization in studies of the Higgs boson and the top
quark.   For studies of an unknown new particle, this tool is even
more powerful.
By measuring  the pair-production rate for the various beam
polarizations, we can directly extract the 
quantum numbers of the particle  under 
the electromagnetic and weak interactions.  We will see illustrations
of the power of beam polarization in the examples discussed below.

If a new particle is discovered or suggested, the energy of the ILC
can be extended to reach the pair-production threshold.  
In addition, the ILC can schedule an energy  scan near  the pair production
threshold, to obtain a very accurate measurement of the particle mass
and quantum numbers.

There is a large literature on ILC searches for new particles,
reviewed in Refs.~\cite{SnowmassBSM,Gudi}.  In this section, we will review a
few especially important examples.  The specific  examples will be taken from
models with supersymmetry, the proposed fundamental symmetry linking
fermions and bosons.  However, the impressive capabilities of the ILC
for new particle searches that these examples illustrate apply more
generally.
   We will also review another
precision probe for new physics, the precision study of two-fermion
pair production.

\subsection{Hidden dark matter}

We have noted already that one of the most important questions in
particle
physics is the particle identity of the dark matter of the universe.
Dark matter particles are difficult to observe at colliders.  They are
neutral and weakly interacting and thus make no signals in collider
detectors.  But  if we are to understand dark matter,  it will be
extremely important to produce dark matter particles in a controlled
environment, to measure their quantum numbers and couplings.  Only
then can we produce a constrained, testable theory of dark matter
production in the early universe~\cite{Baltz}. 

The most powerful strategy for studying dark matter at the LHC is to
produce heavy, strongly interacting particles that decay to dark
matter particles.   The dark matter particles would be invisible to
the LHC detectors, but the visible decay products that accompany them
would give clues to their  properties.  Unfortunately, these heavier
precursor particles have not yet been discovered, and they might in
fact be too heavy to produce at the LHC.   

In many models of dark matter, there is an electrically charged
particle that can decay to the neutral dark matter particle.   If the
charged and neutral particles are well separated in mass, the visible
particles emitted in the decay can be observed at the LHC.   Using
this strategy, the LHC experiments have excluded supersymmetric
partners of the $W$ and $Z$ bosons with masses as high as
700\,GeV in the easiest cases~\cite{ATLASEW,CMSEW}.
However, many interesting cases are much more difficult for the LHC 
experiments.  In particular, it is often true that the lightest neutral particle in these
models has a small annihilation cross section, leading to too many
dark matter particles surviving in the universe today.   To obtain the observed
amount of dark matter, the model should have ``coannihilation'', the
simultaneous annihilation of charged and neutral states.  This
requires that the dark matter particle and its charged partner have a
mass difference of 20~GeV or less.  This gives an example of what
is called a ``compressed'' spectrum in the LHC literature.  The emitted
visible decay products are too soft to pass the triggers of the LHC
detectors, and the process of particle production and decay, which may
occur at a high rate, becomes unobservable.

 A second strategy is to
observe reactions with an initial state radiation gluon recoiling
against invisible particles.  This method allows discovery of the pair-production
of invisible particles.  Most studies for LHC have assumed that dark
matter particles are produced by a pointlike contact interaction. The
pointlike coupling leads to gluon radiation with large transverse
momentum that can be used as a signature~\cite{Whiteson}.  However,
for models,
including supersymmetric models, in which 
dark matter is produced more conventionally by electroweak
interactions,
 this method becomes more difficult.  The spectrum of initial state
 radiation is similar to that in the Drell-Yan process. There is a
large background from production of a $Z$ boson which then decays to
neutrinos,
 and from
$W$ production with a final-state lepton unobserved.    The systematic
error from subtracting this background dominates the measurement.   The
result is that dark matter particles can only 
be discovered  for masses of
100--200\,GeV, depending on the electroweak quantum numbers, even with the 
High-Luminosity LHC~\cite{LHCDM}.   A discovery would  indicate that a new,
invisible particle was produced but would tell us little about its properties.

At the ILC, we can search for production of invisible particles with
initial 
state radiation in a way that addresses these issues.
 Because there is no
large strong interaction background, the ILC experiments need no
triggers and can detect emitted particles with energies below 1\,GeV.
Initial-state radiation is present as photons rather than gluons, so
the background rates are much smaller and also precisely calculable.
The background depends in a known way on beam polarization; this
effect can be used to minimize the background and also to measure its influence.
The sensitivity of the search for dark matter at the ILC through a
search that only relies on  initial state
radiation has been studied in Ref.~\cite{Maxim}.  Any type of dark matter
that annihilates to $\ee$  can be discovered with this technique
(for high enough collider energy), even  if this annihilation channel is
only a few percent of the total annihilation rate.

\begin{figure}
\begin{center}
\includegraphics[width=0.6\hsize]{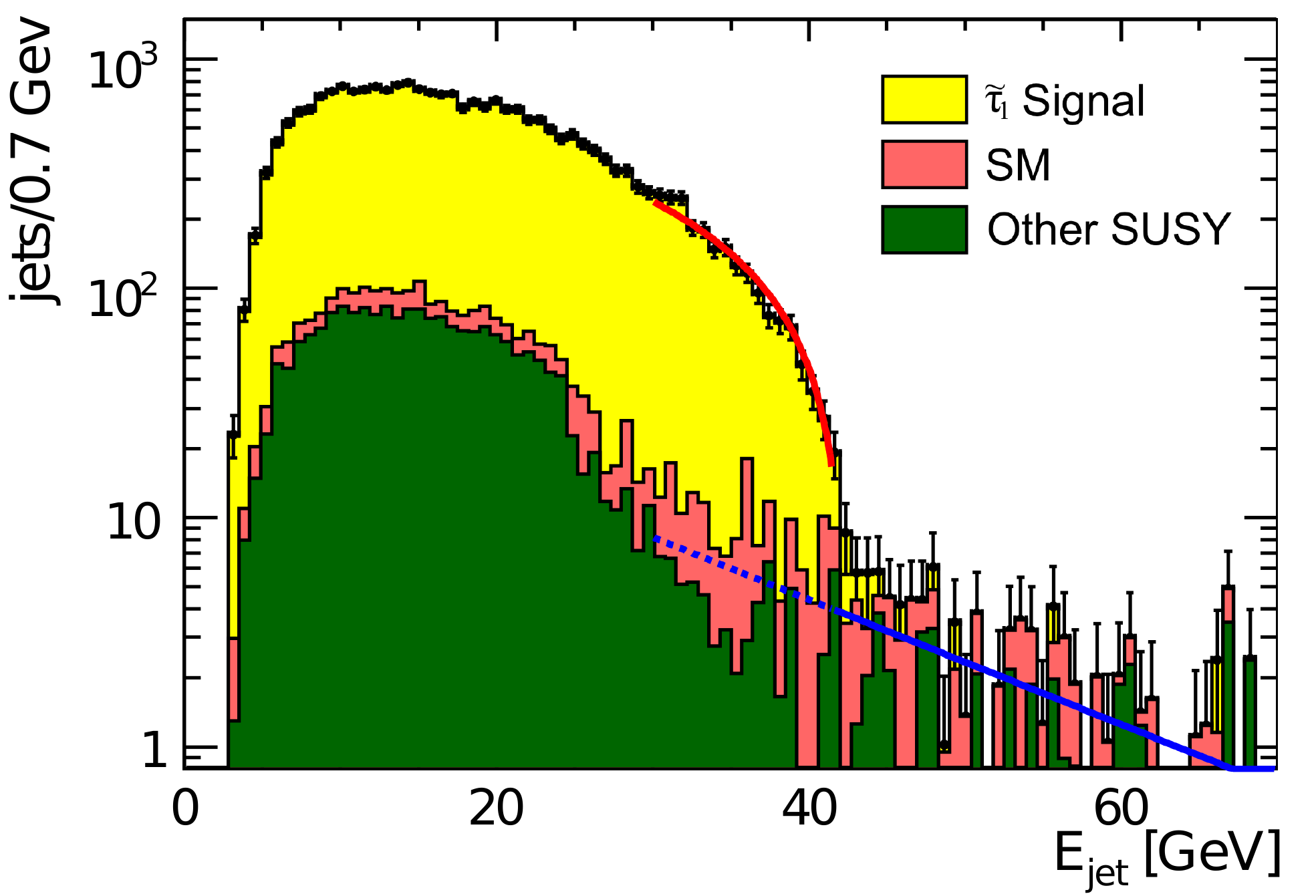}
\end{center}
\caption{Energy distribution of hadronic decay products of the tau
  lepton in events selected for $\s\tau^+\s\tau^-$ pair production at
  the ILC,
  from Ref.~\cite{stau}.}
\label{fig:stau}
\end{figure}

The ILC experiments expect not only to observe  initial state
radiation but also to resolve soft particles produced in decays.   For
example, Ref.~\cite{stau}  considers a model in which the dark
matter particle is the lightest supersymmetry partner $\chi$.   The model is
arranged so that the supersymmetry partner $\s\tau$ of the tau lepton
coannihilates with the dark matter particle $\chi$  to give the observed value
of the cosmic dark matter density.   In the model, the mass of the
$\s\tau $ is 107\,GeV.  The mass difference
between the $\s\tau$ and the $\chi$ is set by the 
coannihilation rate to be 11\,GeV.
Figure~\ref{fig:stau}  shows an ILC simulation of
pair-production of the supersymmetric partner ($\s \tau$)  of the tau
lepton.  The $\s\tau$ decays to a $\chi$ and a 
soft tau lepton, which is observed
in its decay to low-energy $\pi$ mesons. 
The $\s\tau$ pair production signal,
shown in yellow in the figure, stands out clearly above the various
backgrounds.  In the analysis of this model, the masses of
the $\s\tau$ and $\chi$ are determined to a precision of 
200\,MeV and 400\,MeV,
respectively.  The electroweak quantum numbers of the $\s\tau$ are
determined from the production rates with polarized beams.
 By combining this information with other information
available from ILC measurements in this model, the annihilation rate
of the $\chi$ can be determined and 
the cosmic density of $\chi$ dark matter can be 
predicted to 0.2\% accuracy.

\subsection{Hidden Higgsino}

Supersymmetry is an especially attractive principle to extend the
Standard Model.   It gives a framework for the unification of the
coupling constants of the strong, weak, and electromagnetic
interactions and a {\it raison d'\^etre} for the appearance of
fundamental scalar fields such as the Higgs field.    New particles
predicted by supersymmetry have been searched for intensively at the
LHC, though none have been found so far. 
Supersymmetry at the weak interaction scale is often motivated by its
possible role in explaining the form of the Higgs
potential.  For this, some supersymmetric particles must have masses
near the weak interaction mass scale.  The strongest arguments can be
made that the supersymmetric partners of the Higgs boson, called
Higgsinos,
should be  relatively light and accessible to collider
experiments~\cite{naturalHiggsino}.

However, the Higgsinos are especially difficult to discover at the LHC.
They  have all of the problems described for dark matter in the previous
section. They are produced only by electroweak interactions.  Though
there must be both charged and neutral Higgsinos, supersymmetry
predicts that their masses are
naturally
compressed  if the supersymmetric
partners of the weak interaction gauge bosons are heavy.  Thus it is
not surprising that Higgsinos are hardly constrained by LHC data.

\begin{figure}
\begin{center}
\begin{minipage}[c]{0.46\textwidth}
\includegraphics[width=0.99\textwidth]{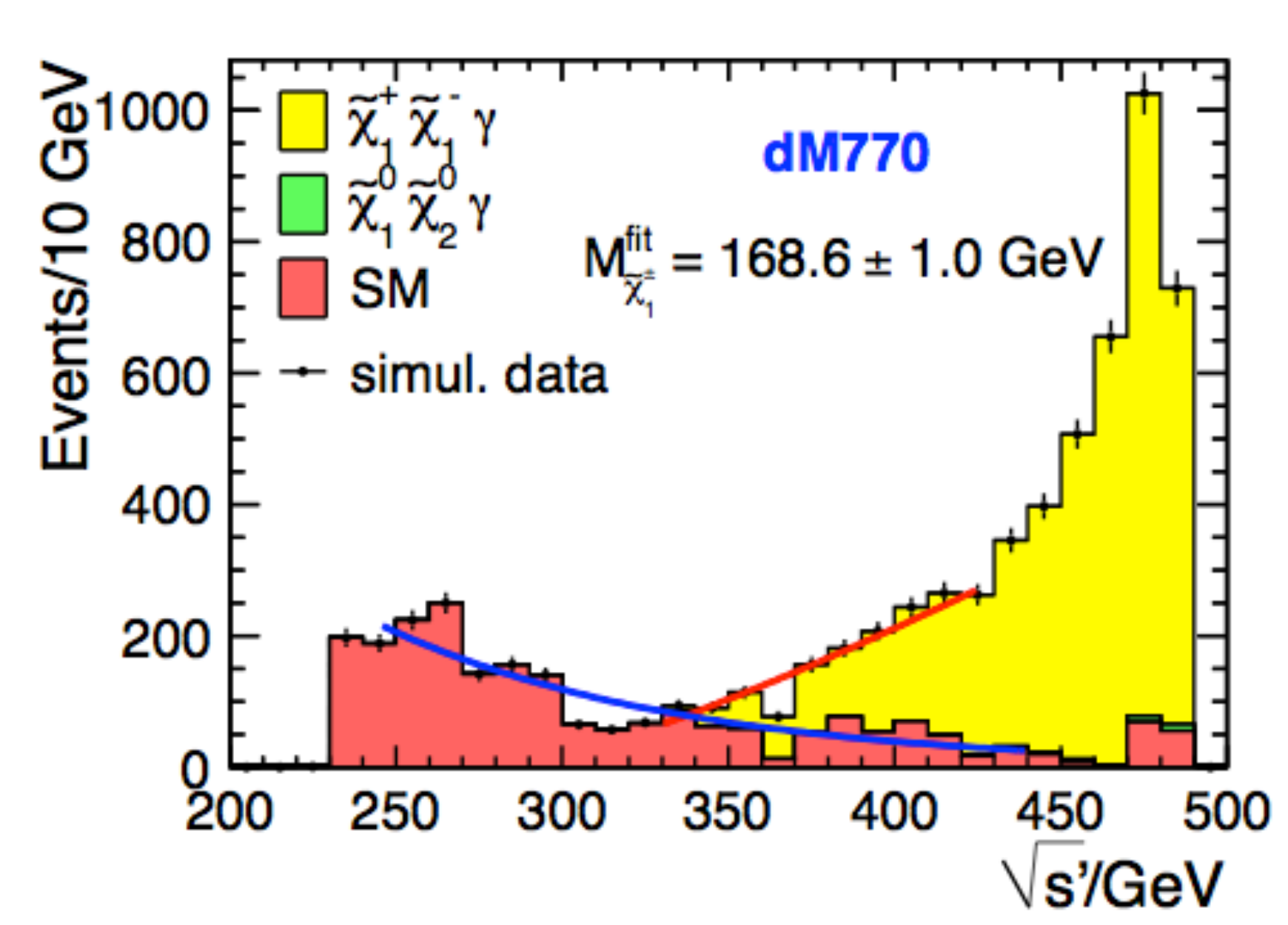}
\end{minipage}
\hspace{0.03\textwidth}
\begin{minipage}[c]{0.46\textwidth}
\includegraphics[width=0.99\textwidth]{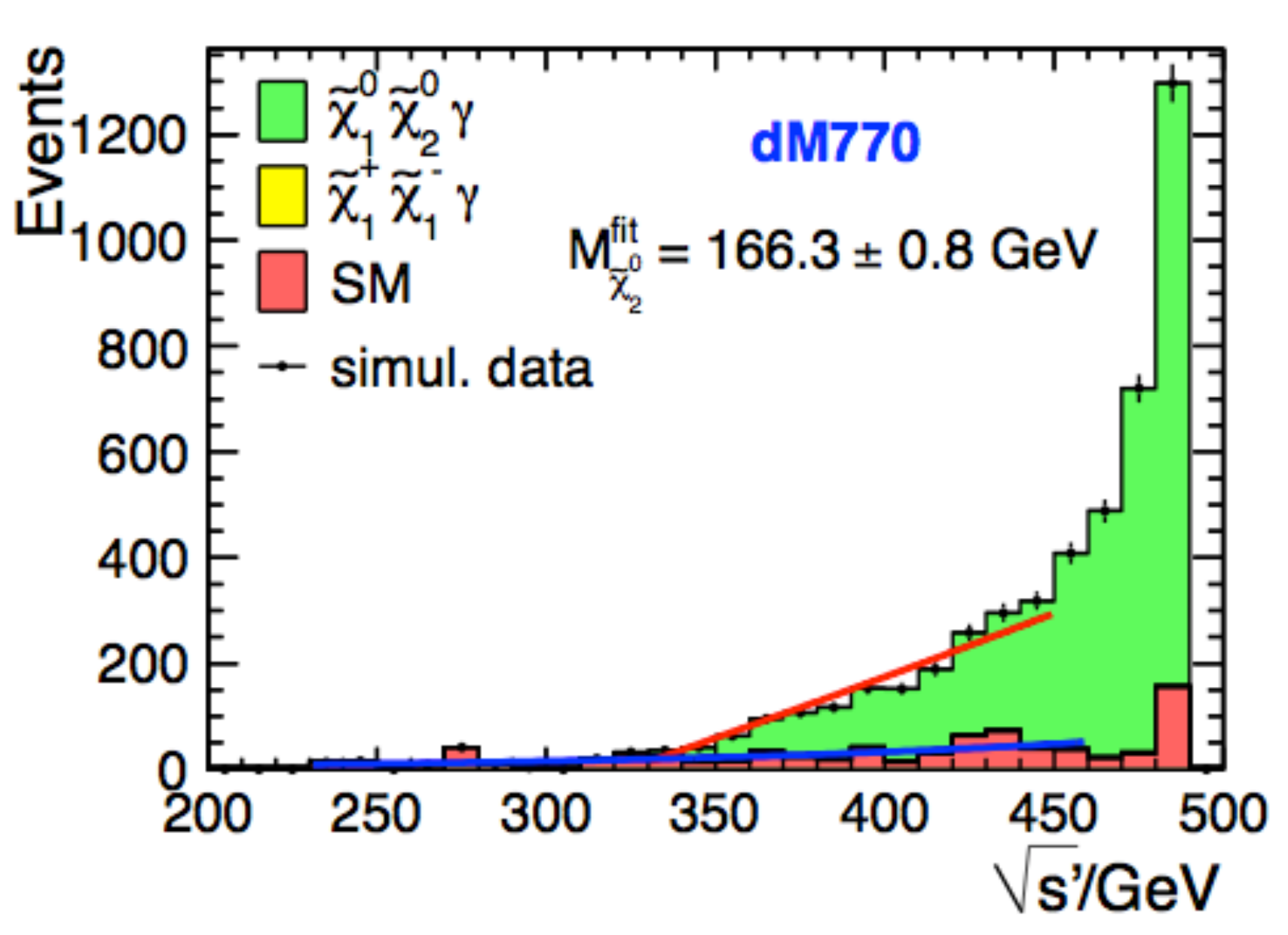}
\end{minipage}
\end{center}
\caption{Distribution of the missing mass for a system recoiling
  against an initial state radiation photon in a model with Higgsino
  production at the ILC,
  from Ref.~\cite{Higgsino}. Left:  events selected for charged Higgsinos.
  Right:  events selected for neutral Higgsinos.}
\label{fig:Higgsino}
\end{figure}

The  observation of Higgsinos at the ILC has been studied 
in Ref.~\cite{Higgsino}.  The more difficult model considered in
this paper contains Higgsinos with masses near 165\,GeV, with a mass
difference of about 1\,GeV between the charged and the lighter neutral
Higgsino.
Nevertheless the signal of Higgsino production from  initial state
radiation photons is substantial, as shown in Fig.~\ref{fig:Higgsino}.
The various Higgsino masses are determined to a precision of  about 1\,GeV.
In addition, it is possible to observe soft  $\pi$  mesons from the
decay of the charged
Higgsino, providing a very sharp determination of the mass difference
between the Higgsino states.  The mass differences provide an estimate
of the masses of the supersymmetric partners of the electroweak gauge
bosons, which are set to several TeV in this model. 

From  the rates for Higgsino production with polarized beams, we will
be able to determine the quantum numbers of these particles.   In this case, the cross
section measurements can confirm that the particles discovered indeed
have the quantum numbers expected for a Higgsino.

\subsection{Hidden pseudoscalar Higgs bosons}

Another context in which the ILC can make important additions to the 
LHC searches for new particle is in the search for additional Higgs
bosons.   We have already discussed the prospects for searching for
such bosons indirectly through shifts in couplings of the known Higgs
bosons to quarks and leptons.  There are also many scenarios in which
additional Higgs bosons are relatively light but difficult to discover
at the LHC.  For example, a model with two Higgs fields contains a 
Higgs boson of odd parity that decays primarily to heavy quarks and
leptons.   This boson can be as light as the known Higgs boson, or
even 
lighter, due to additional global symmetries of the Higgs sector.
Because of its parity, its coupling to the $W$ and $Z$ bosons is
suppressed.
If the mass of this particle  were below 10\,GeV, it would
be seen in $\Upsilon$  decays; above 200\,GeV, it  could be seen
at the LHC in decays to two photons or two gluons.  Between these
limits, the LHC must search for these  particles using electroweak
production and purely leptonic decays, a difficult prospect even for
the high-luminosity era. At the ILC running at 500\,GeV,  this Higgs
boson is produced by
radiation from top or bottom quarks. It can be discovered
straightforwardly in its
dominant
$b \bar b$ and $\tau^+\tau^-$ decay modes~\cite{Kilian:2004pp}.  If
the mass of the odd parity Higgs boson is less than 62\,GeV, it can
also be discovered as an exotic decay product of the known Higgs boson~\cite{TaoLiu}.

\subsection{Two-fermion processes}

The ILC will also be able to search for new gauge bosons, making use
of its capability for precision
measurement of the basic  two-fermion processes $\ee\to \ell^+\ell^-$
and $\ee\to q\bar q$.  A new  neutral gauge boson $Z'$ will perturb
the cross sections predicted in the Standard  Model through
interference with the production diagrams involving the  $\gamma$ and
$Z$.  The Standard Model expectation is understood theoretically at
the 0.1\% level, so these measurements are sensitive to $Z'$ bosons
well above the collider center of mass energy.

\begin{figure}
\begin{center}
\includegraphics[width=1.0\hsize]{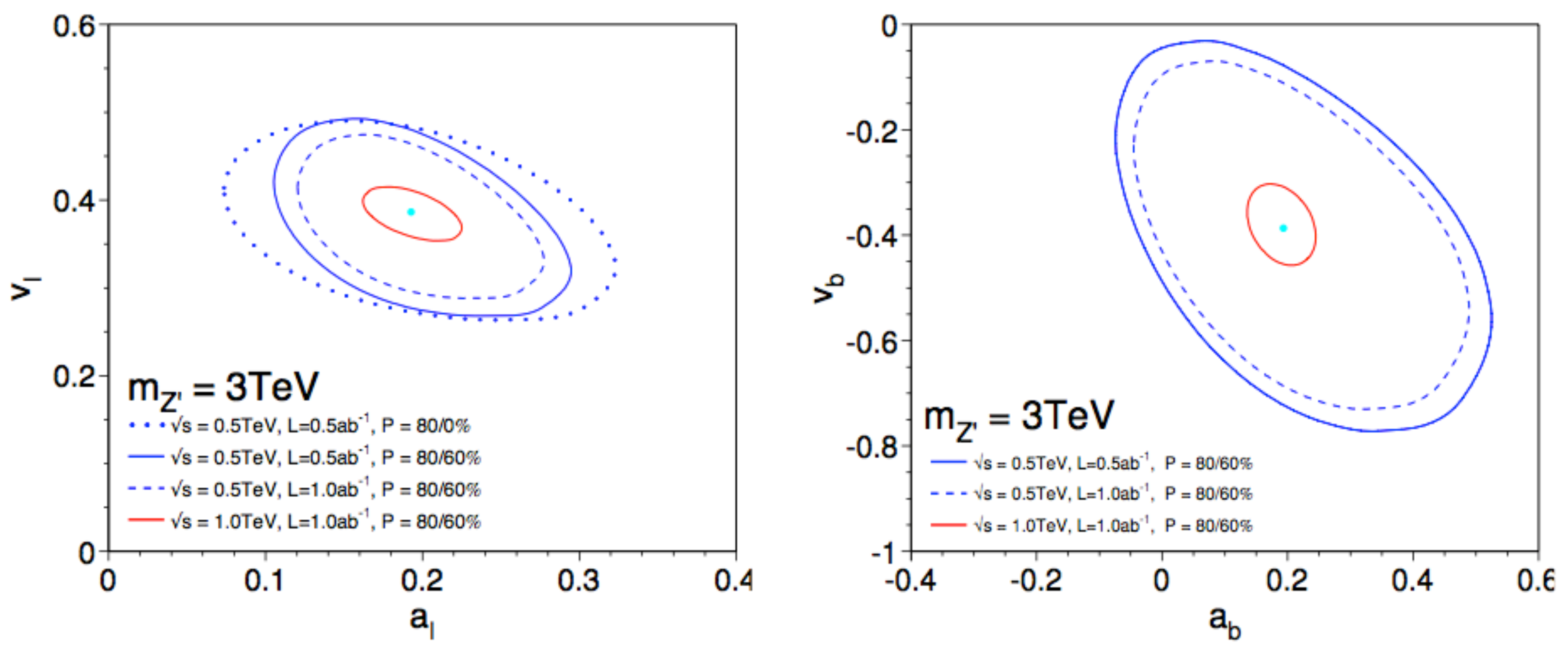}
\end{center}
\caption{Determination of the vector and axial couplings of a $Z'$
  resonance to leptons (left panel) and $b$ quarks (right panel) through measurement of $\ee\to
  \ell^+\ell^-$ and $\ee\to b\bar b$ at the ILC, from Ref.~\cite{ILCTDR}. }
\label{fig:Zprime}
\end{figure}

New $Z'$ bosons  appear, for example, in models of the grand
unification of the strong, weak, and electromagnetic interactions.
The gauge group that unifies the known interactions may include other,
new, interactions as well. 
Figure~\ref{fig:Zprime} shows an analysis of a model with a $Z'$ boson with the
quantum numbers expected from 
the grand unification group $SO(10)$.  In the example studied, the boson
is assumed to have a  mass of 3~TeV.  Such a boson has been
searched for at the LHC as a resonance in lepton pair production and
excluded for masses up to 2.6~TeV~\cite{ATLASZprime}.   If the
resonance is present at a higher value of the mass, the figure
illustrates that the ILC will give significant information about its
pattern of couplings.
 If the resonance is not present, the ILC will put a
lower bound on its mass at 7~TeV, comparable to the projected
LHC bound of 6~TeV.  Running of the ILC at 1000~GeV would roughly
double its search reach~\cite{ILCTDR}. 

If a $Z'$ is indeed present in the region that will be explored by the
LHC and the ILC, there will be  impressive synergy between the
measurements at the two colliders.  The LHC experiments will observe
the $Z'$ directly as a resonance and measure its mass accurately.  The
ILC will then measure the couplings of the $Z'$ to each individual
quark and lepton species, taking advantage of beam polarization to
measure both the  left-handed and right-handed couplings in each case.
With this information, one can fully identify the gauge boson and find
its place in an extended gauge theory of nature.

Other effects can also perturb the two-fermion processes.  If quarks
and leptons are composite, the first manifestation of this will be the
perturbation of two-fermion processes by the effects of higher-dimension
operators.   This effect can be searched for at the LHC, but it
requires careful calibration of quark jet signals at the highest
energies.  In addition, limits from the LHC  are model-dependent because many operators can
potentially contribute.  For this reason, the best current
constraints on
quark and lepton compositeness still come from the  data from the
$\ee$ collider LEP,
putting limits on the compositeness scale at about 10~TeV.  The ILC,
with higher energy, luminosity, and intrinsic precision and also the
capability for  
electron and positron beam polarization, should
improve these constraints by an order of magnitude in the
compositeness scale~\cite{ILCTDR}.

\section{Conclusion}

In this report, we have surveyed the major elements of the ILC physics
program.  We have reviewed the ILC capabilities to search for new
particles and interactions through precision studies of the Higgs
boson and the top quark, and we have reviewed the capabilities of the ILC
to carry out direct searches for possible new particles.

The discovery of the Higgs boson at the LHC has completed the
construction of a Standard Model for particle physics.  That model is
potentially complete and internally consistent. Howver, it is also inadequate
to answer the many open questions that remain in particle
physics.   If these questions have answers, the Standard Model must
break down. Researchers in our
community are asking where and how that breakdown can be discovered.

The ILC offers new avenues to address that question experimentally.
It offers new, precise, unambigous information on the two elementary
particles most closely connected to our questions about the Standard
Model, the top quark and the Higgs boson itself.

Though the LHC has carried out many searches for new particles, there
are gaps in those searches reflecting the difficulty of
experimentation at hadron colliders. The ILC will bring new 
capabilities that will allow crucial searches
for a variety of well-motived new physics scenarios such as
supersymmetry, Higgs compositeness, new gauge bosons, additional Higgs
bosons, and particles connected with cosmic dark matter that may not
be possible to see by any other means.

From these arguments, we know today that the ILC has the potential to 
make major
contributions to particle physics.   As the LHC winds down its program
fifteen years from now, the ILC will become the world's most important
source of new
information on the issues that surround the Standard Model.  If new
physics beyond the Standard Model is discovered in that interval,
through results from the LHC, from dark matter detection, from
cosmology, 
or from other
sources, so much the better. 

Finally, we note that, although the estimates of performance of the
ILC experiments given in this report are done as realistically as
possible at this stage, it is another question to extract these levels
of performance from a running experiment.  The ATLAS and CMS
collaborations have now met and even exceeded the performance they
projected before the startup of the LHC. But this accomplishment took
the hard work of many people, in optimizing and then constructing the
detectors, understanding the actual environment provided by the
accelerator, producing precise calculations and simulations, and
collecting and analyzing the data in this context. For the ILC, a
similar effort will be needed to carry out the program we have
outlined here. 
We hope that readers of this paper will join us in this endeavor 
to realize the promise offered by the ILC.

\bigskip

\Acknowledgements

We thank James Brau, Roberto Contino, and Junping Tian for their assistance in
preparing this report.  We are grateful to many other members of the
Linear Collider community for discussions of the issues reviewed here.

\bigskip

\appendix

\section{Appendix: Table of ILC projected uncertainties}

In this appendix, we list the current projections from  the ILC detector
groups for the expected accuracy with which the most important
physics parameters constrained by the ILC will be measured.  We
recommend that these numbers be used in discussions of the ILC physics
prospects and in comparisons of the ILC with other proposed
facilities.

Projected accuracies for the ILC depend on the run plan assumed.
  Following the report of the ILC Parameters Joint Working 
Group ~\cite{ParameterGroup}, we assume the following 
scenario:   In its initial phase, the ILC would accumulate
500~fb$^{-1}$ at 500~GeV, then 200~fb$^{-1}$ at 350~GeV, then
500~fb$^{-1}$ at 250~GeV.  After a luminosity upgrade, the ILC would
accumulate an additional 3500~fb$^{-1}$ at 500~GeV, then 1500~fb$^{-1}$ at 250~GeV.
Based on the accelerator design described in the
 ILC TDR~\cite{ILCTDRmachine}, this program would
require 
8.1 calendar years, including a realistic
start-up profile, to complete the initial phase.  It would require a  total of 20.2
calendar years to complete the whole program, including the downtime needed
for the upgrade.   The full ILC data set would then include
2~ab$^{-1}$ at 250~GeV, 200~fb$^{-1}$  at 350~GeV and 4~ab$^{-1}$  at
500~GeV.    More details and some alternative scenarios are
given 
in \cite{ParameterGroup}.

Table~\ref{tab:resultsone} gives the corresponding projections for
the uncertainties in physics parameters.   The numbers listed are
obtained
 from full-simulation analyses using
the ILD and SiD detector models described in Ref.~\cite{ILCTDRdetectors}.   
These are  Geant4-based
 simulations with detailed  detector designs,
which have in many cases been confirmed by
test beam measurements on detector prototypes.
For
each number, we have given a reference in which the method is 
described.  The actual number given may reflect more recent
improvements in the analysis~\cite{newrefs}.   The uncertainties include both
expected statistical and systematic errors.

The estimated uncertainties for the Higgs boson couplings are based on
a fit to ILC observables in which all individual couplings (including
the loop-induced couplings to $gg$ and $\gamma\gamma$) are varied
independently.  The total width of the Higgs boson is also taken as an
independent variable, to account for exotic Higgs decays not
constrained by any direct measurement.  (Higgs decay to invisible
states is directly observed using the $hZ$ production process.) 
  The constraints on the Higgs boson derived from
this analysis are completely model-independent.  

The second line for $g(h\gamma\gamma)$ assumes that the ILC data are
combined with an LHC measurement of the ratio of branching ratios 
$\Gamma(h\to \gamma\gamma)/\Gamma(h\to ZZ ^*)$ to 2\% accuracy, as
projected by the ATLAS collaboration for the High-Luminosity
LHC~\cite{ATLAS}.
This is the only place where combination with projected LHC results
significantly improves the model-independent ILC results.

For comparison with results from hadron colliders, where a
model-independent analysis is not possible, the 2013 Snowmass study~\cite{SnowHiggsReport}
suggested
a fit to observables in which one adds the model assumptions that 
$g(hc\bar c)/g(ht\bar t)$ and   $g(h\mu\mu)/g(h \tau\tau)$ have
their Standard Model values and that the Higgs boson has no invisible or
exotic decays.   The results of that analysis are given in
Table~\ref{tab:resultstwo}. 

 Results from the analysis of measurements of the top quark threshold
are dominated by theoretical systematic errors from the theoretcial calculation of
the threshold cross section shape.  These errors are estimated
conservatively based on a new calculation of this cross section at the 
N$^3$LO level~\cite{Beneke,Beneketwo}.   Measurements at any $\ee$ collider should show these 
same uncertainties.   The statistical errors from the ILC program are
quoted in the main text.

The estimated uncertainties for the top quark electroweak couplings are analyzed
in the following way~\cite{Roman}: First a fit is done with the four
chirality-conserving couplings
$g_L^\gamma, g_R^\gamma, g_L^Z, g_R^Z$ taken to be independent
parameters and the chirality flip couplings taken to be zero.  Then a
fit is done with the chirality conserving parameters taken at their
Standard Model   values and the two  CP-conserving chirality-flip
parameters
$F_2^\gamma$, $F_2^Z$ taken as independent free parameters.    For the
high-luminosity estimates, we have conservatively added a 0.5\%
systematic uncertainty.

The limits on dark matter production are based on an effective field
theory analysis as described in Ref.~\cite{Maxim}. Dark matter pair production
is represented by a contact interaction with the scale $\Lambda$; the labels D5 and D8 refer to
two possible spin structures.  We emphasize that the
effective field theory approximation is accurate  in this analysis, while
it is questionable in similar analyses for hadron colliders.   The
quoted limits are based on a full-simulation study described in Ref.~\cite{JDM}. 

\begin{table}[h] \begin{center}
\begin{tabular}{lc|c|c|lr}
Topic          &  Parameter   & Initial Phase & Full Data Set &  units
&  ref.  
\\  \hline 
Higgs          &   $ m_h $      &   25    & 15  &  MeV & \cite{Li}
\\
                    &   $   g(hZZ)   $    
                    &   0.58   &  0.31  &   \% &
                    \cite{SnowmassHiggs} \\
                   &     $   g(hWW)     $ &  
                   0.81 &  0.42       &    \% &
                    \cite{SnowmassHiggs} \\
                   &     $   g(hb\bar b)   $ & 
                  1.5 &  0.7   &    \% &
                    \cite{SnowmassHiggs} \\  
                   &    $    g(h g g)   $ & 
                  2.3 & 1.0     &      \% &
                    \cite{SnowmassHiggs} \\
                   &    $    g(h \gamma \gamma)   $  &  7.8  &   3.4 &\% &
                    \cite{SnowmassHiggs} \\
                 &   &   1.2  & 1.0&   \%,
                  w. LHC results & \cite{MEPHiggs}\\
                   &      $ g(h \tau\tau)   $ & 
                   1.9 &   0.9 &  \% &
                    \cite{SnowmassHiggs} \\
                   &    $   g(h c\bar c)    $   &   
                2.7 &   1.2 & \% &
                    \cite{SnowmassHiggs} \\
                   &    $   g(h t\bar t)    $  &  18 &   6.3     & \%,
                direct     &
                    \cite{SnowmassHiggs} \\
                 &    &  20 &   20     & \%, 
                 $t\bar t$ threshold  &
                    \cite{Horiguchi} \\
                 &    $   g(h\mu\mu)    $   &     20    &  9.2    &  \% &
                    \cite{SnowmassHiggs} \\
                   &      $ g(hhh)     $         & 77 &  27     & \% &
                    \cite{SnowmassHiggs} \\
                    & $ \Gamma_{tot}$  & 3.8 &
                 1.8  & \% &
                    \cite{SnowmassHiggs} \\
                   & $ \Gamma_{invis}$ & 0.54&
                 0.29    & \%,   95\% conf. limit  &
                    \cite{SnowmassHiggs} \\
\hline
Top         &    $  m_t $        &  50 & 50  &  MeV ($m_t$(1S))  &
                                   \cite{Seidel:2013sqa} \\
        & $ \Gamma_t $  &60   & 60  &  MeV &  \cite{Horiguchi}   \\ 
                &    $ g_L^\gamma   $ & 0.8  &
               0.6  & \% & \cite{Roman}\\
                          &    $ g_R^\gamma   $ & 0.8  &
               0.6  & \% & \cite{Roman}\\
    &    $ g_L^Z   $ & 1.0 & 0.6  & \% &  \cite{Roman}\\
                &    $  g_R^Z   $ & 2.5 &
               1.0   & \% &  \cite{Roman}\\
                 &  $ F_2^\gamma$  & 0.001 & 0.001  & absolute  &  \cite{Roman}\\
                 &  $ F_2^Z$  &  0.002  & 0.002  &absolute &  \cite{Roman}\\
       \hline
$W$       &  $ m_W   $   &     2.8     &   2.4  & MeV   &\cite{Wmass} \\
         &   $ g^Z_1     $    &        $    8.5\times 10^{-4}     $   &
        $  6\times 10^{-4}  $ &   absolute  & \cite{List} \\
  &  $\kappa_{\gamma}   $   &   $  9.2\times 10^{-4}     $    &    $   7 \times 10^{-4}  $   &
  absolute &  \cite{List} \\ 
   &  $\lambda_{\gamma}    $  &  $ 7 \times 10^{-4}    $    &    $   2.5 \times 10^{-4}   $&  absolute &  \cite{List} \\ 
\hline
Dark Matter     &  EFT $\Lambda$: D5  &  2.3 & 3.0 & TeV,
90\% conf. limit  & \cite{JDM}\\ 
&  EFT $\Lambda$:  D8  &  2.2 & 2.8 & TeV,
90\% conf. limit  & \cite{JDM} \\ \hline
\end{tabular}

\caption{Projected accuracies of measurements of Standard Model
  parameters  at the two stages of the ILC program proposed in the
  report
of the ILC Parameters Joint Working Group~\cite{ParameterGroup}.  This program has  an
  initial phase with 500~fb$^{-1}$
  at 500~GeV,   200~fb$^{-1}$  at 350~GeV,  and  500~fb$^{-1}$
  at 250~GeV,  and a luminosity-upgraded phase with an additional 3500~fb$^{-1}$
  at 500~GeV and 1500~fb$^{-1}$
  at 250~GeV.  Initial state polarizations are taken according to the
  prescriptions of~\cite{ParameterGroup}.  Uncertainties are
  listed as $1\sigma$ errors (except where indicated),
  computed cumulatively at each stage of the program.  These estimated
  errors include
  both statistical uncertainties and theoretical and experimental systematic
  uncertainties. Except where indicated, errors in 
  percent (\%)  are fractional uncertainties
  relative to the Standard Model values. More specific information for
  the sets of measurements is given in the text. For each measurement, a
  reference describing the technique is given. }
\label{tab:resultsone}
\end{center}
\end{table}

\begin{table}[h] \begin{center}
\begin{tabular}{lc|c|c|l}
Topic          &  Parameter   & Initial Phase & Full Data Set &
\\  \hline 
Higgs          &   $   g(hZZ)   $    
                    &   0.37   &  0.2  &   \% \\
                     &     $   g(hWW)     $ &  
                   0.51 &  0.24      &    \% \\
                   &     $   g(hb\bar b)   $ & 
                  1.1&  0.49   &    \% \\  
                   &    $    g(h g g)   $ & 
                  2.1 & 0.95      &      \% \\
                   &    $    g(h \gamma \gamma)   $  & 7.7  &   3.4 &\% \\
                  &      $ g(h \tau\tau)  , g(\mu\mu) $ & 
                   1.5 &   0.73 &  \%\\
                   &    $   g(h c\bar c), g(ht\bar t)    $   &   
                2.5 &   1.1  & \% \\
                    & $ \Gamma_{tot}$  & 1.8  &
                  0.96  & \% \\   \hline
\end{tabular}

\caption{Projected accuracies of measurements of Higgs boson couplings
  at the two  stages of the ILC program, from  the
  model-dependent fit used in the Snowmass 2013
  study~\cite{SnowHiggsReport}. The analysis is as described in
  \cite{SnowmassHiggs}.  The ILC run plan assumed is the same as in 
  Table~\ref{tab:resultsone}. }
\label{tab:resultstwo}
\end{center}
\end{table}



\begin{thebibliography}{99}
\bibitem{newrefs}
 More information about the updated analyses can be 
found in  Ref.~\cite{ParameterGroup} and in workshop talks
 cited there.
\end{thebibliography}
\end{document}